\documentclass[11pt, aps,prd, preprintnumbers, floatfix, superscriptaddress,pacs, showpacs]{revtex4}

\usepackage[dvips]{graphics}
\usepackage{amsmath,amsfonts,bm,color}
\usepackage{epsfig,bm}
\usepackage{graphicx,comment,epsf}
\begin{document}

\preprint{JLAB-THY-07-708}

\affiliation{Thomas Jefferson National Accelerator Facility,
             Newport News, VA 23606, USA}
\affiliation{Physics Department, Louisiana State University,
             Baton Rouge, LA 70803, USA}
\affiliation{Physics Department, Old Dominion University, Norfolk,
             VA 23529, USA}

\author{H.~R.~Grigoryan}
\affiliation{Thomas Jefferson National Accelerator Facility,
             Newport News, VA 23606, USA}
\affiliation{Physics Department, Louisiana State University,
             Baton Rouge, LA 70803, USA}
\author{A.~V.~Radyushkin}
\affiliation{Thomas Jefferson National Accelerator Facility,
              Newport News, VA 23606, USA}
\affiliation{Physics Department, Old Dominion University, Norfolk,
             VA 23529, USA}
\affiliation{Laboratory of Theoretical Physics, JINR, Dubna, Russian
             Federation}

\title{Pion Form Factor in Chiral Limit of Hard-Wall   AdS/QCD Model}

\begin{abstract}
We describe a formalism to calculate form factor and 
charge density distribution of the pion in the chiral limit 
using the holographic dual model of QCD with hard-wall cutoff.
We introduce two conjugate pion wave functions
and present analytic expressions for these functions 
and for the pion form factor.  
They allow to relate such observables 
as the pion decay constant and the pion charge electric radius  
to  the values of chiral condensate and hard-wall cutoff scale.  
The evolution of the pion form factor to large values
of the momentum transfer is discussed,
and results are compared to existing experimental data.

\end{abstract}

 \maketitle

\section{Introduction}

During the last few years applications of gauge/gravity
duality \cite{Maldacena:1997re} to hadronic physics
attracted a lot of attention, and  various holographic dual models of
QCD were proposed in the literature (see,
e.g.,
\cite{Polchinski:2002jw,Boschi-Filho:2002vd,Brodsky:2003px,Sakai:2004cn,Erlich:2005qh,Erlich:2006hq,DaRold:2005zs,Karch:2006pv,Csaki:2006ji,Hambye:2005up,Hirn:2005nr,Ghoroku:2005vt,Brodsky:2006uq,Evans:2006dj,Casero:2007ae,Gursoy:2007cb,Gursoy:2007er,Bergman:2007pm,Erdmenger:2007vj,Dhar:2007bz}). 
These models were able to
incorporate such essential properties of QCD as confinement 
and chiral symmetry breaking, and also to reproduce many
of the static hadronic observables (decay constants, masses), 
with values rather  close to the experimental  ones.
Amongst the  dual models,  a special class is the so-called ``bottom-up'' approaches (see, e.g.,
\cite{Erlich:2005qh,Erlich:2006hq,DaRold:2005zs,Karch:2006pv}), 
the goal of which is to reproduce known properties
of QCD by choosing an appropriate theory in the 5-dimensional (5D) AdS bulk.
 Within the framework of the  AdS/QCD models, by modifying
the theory in the bulk one  may try to explain/fit
experimental results in different sectors of
QCD.

In the present paper, we will be interested in the  hard-wall AdS/QCD model 
\cite{Erlich:2005qh,Erlich:2006hq,DaRold:2005zs},  where 
the confinement is modeled by
sharp  cutting off the AdS space along the extra fifth dimension 
at a wall located at some finite distance  $z = z_0$.
In the framework of this hard-wall model,  it is  possible to find  form
 factors and wave functions of vector
mesons (see, e.g.,  \cite{Grigoryan:2007vg}).
 To reproduce the general features of the spectrum for
the higher states (``linear confinement''), a  soft-wall model was
proposed in \cite{Karch:2006pv}.  The $\rho$-meson form factors
 for  this  
model were calculated in Ref.~\cite{Grigoryan:2007my}. 

In general, the vector sector is less sensitive to the infrared (IR)
effects, since this symmetry is not broken in
QCD. However, the axial-vector sector appears to be
very sensitive to the particular way the chiral symmetry is broken or, in
other words,  to the bulk content and the shape of the  IR wall \cite{Karch:2006pv}.

In this respect, one of the interesting objects
to study in the holographic dual models of QCD is the pion. The
properties of the pion were studied in various holographic approaches,
(see e.g. Refs.~\cite{Sakai:2004cn, Erlich:2005qh,
DaRold:2005zs, Hirn:2005nr, Evans:2006dj, Ghoroku:2005vt,Casero:2007ae,Bergman:2007pm,Dhar:2007bz,Brodsky:2006uq,Radyushkin:2006iz}).
 In particular, the approach of
Ref.~\cite{Erlich:2005qh} (see   also recent papers  \cite{Casero:2007ae,Bergman:2007pm,Dhar:2007bz}) 
 managed to reproduce
 the (Gell-Mann--Oakes--Renner) relation $ m^2_{\pi} \sim m_q $ between the
quark mass $ m_q $ and mass of the pion $ m_{\pi} $ and also
the $ g_{\rho \pi\pi} $ coupling (the coupling between
$\rho$ meson and two pions).
In Ref.~\cite{DaRold:2005zs}, the solution of the pion wave-function equation
was  explicitly written for the $m_q=0$ limit.

In this paper, working in the framework of the model
proposed in \cite{Erlich:2005qh}  (hard-wall model), we
describe  a formalism to calculate the form factor and wave
functions (and also the density function) of the pion. 
Since the  fits of Ref.~\cite{Erlich:2005qh} give 
a very small $m_q \sim 2\,$MeV value for 
the explicit chiral symmetry breaking parameter $m_q$,
we consider only  the chiral limit $m_q=0$
of the hard-wall holographic dual model of two-flavor QCD.
Resorting to  the chiral limit allows us to 
utilize one of the main advantages of AdS/QCD -
the possibility to work with  explicit analytic 
solutions of the basic equations of motion.
Expressing the pion form factor in terms of these 
solutions, we are able, in particular,  to  extract and analyze the behavior
of the pion electric radius in  various regions
of the  holographic parameters space. On the numerical side,
we come to 
the conclusion  that the radius of the
pion is smaller than  what is known from  experiment.
However, we suggest that, as in case of the radius of the
$\rho$ meson,    smoothing the IR wall may increase the pion radius.

In our analysis, we  introduce and systematically use
  two types of holographic
wave functions  $\Phi (z)$ and $\Psi (z)$, which are 
conjugate to each other and  basically   similar to the
analogous objects introduced in our papers 
\cite{Grigoryan:2007vg,Grigoryan:2007my},
where we studied  vector mesons.

The paper is organized in the following way. 
We start with recalling, in Section II, 
the basics of the hard-wall model and  some results obtained in
Ref.~\cite{Erlich:2005qh}, in particular, the form of 
the relevant action, the eigenvalue equations for bound
states and their solutions. In  Section III, we describe 
 a formalism for  calculating  the pion form factor  and express
it in terms of the two wave  functions mentioned above. 
In Section IV, we discuss the relation of our AdS/QCD
results  to  experimental data.  We express  the values of
the  pion  decay constant and the pion charge radius in
terms of the fundamental parameters of the theory and 
study their behavior in different regions of the parametric
space. At the end,  we study   the  behavior of the pion form factor at large momentum transfer. Finally, we
summarize the paper.

\section{Preliminaries}

In the holographic model of hadrons,  QCD resonances correspond to Kaluza-Klein (KK) excitations in the sliced
AdS$_5$ background. In particular, vector mesons correspond to the KK modes of transverse vector gauge field in
this background. Since the gauge symmetry in the vector sector of the H-model is not broken, the longitudinal
component of the vector gauge field is unphysical, and only  transverse components correspond to physical mesons.
Similarly, the axial-vector mesons are the modes of the transverse part of the axial-vector gauge field.  However,
because the axial-vector gauge symmetry is broken in the 5D background, the longitudinal components have physical
meaning and are related to the pion field. This should be taken into account if 
we want to treat the  pion  in a
consistent way.

\subsection{Action and Equations of Motion}

The standard prescription of the holographic model is that there is a correspondence between the 4D vector and
axial-vector currents and the corresponding 5D gauge fields:
\begin{align}
J^a_{V\mu}(x) &= \bar{q}(x)\gamma_{\mu}t^a q(x) \rightarrow V^a_{\mu}(x,z) \\ \nonumber J^a_{A\mu}(x) &=
\bar{q}(x)\gamma_{\mu}\gamma_5 t^a q(x) \rightarrow A^a_{\mu}(x,z) \ ,
\end{align}
where $ t^a = \sigma^a/2 $, ($ a = 1,2,3 $ and $ \sigma^a $ are usual Pauli matrices).

In general, one can write $ A = A_{\perp} + A_{\parallel} $, where $ A_{\perp} $ and $ A_{\parallel} $ are
transverse and longitudinal components of the axial-vector field. The spontaneous symmetry breaking causes $
A_{\parallel} $ to be physical and associated with the Goldstone boson, pion in this  case. The longitudinal
component may be written in the form: $ A^a_{M \parallel}(x,z) = \partial_M \psi^a(x,z) $. Then $ \psi^a(x,z) $
corresponds to the pion field. Physics of the axial-vector and 
pseudoscalar sectors is described by the action
\begin{align}
\label{AdS} S^{A}_{\rm AdS} &= {\rm Tr}  \int d^4x~dz~\left[\frac{1}{z^3}(D^{M}X)^{\dagger}(D_{M}X) +
\frac{3}{z^5} X^{\dagger}X - \frac{1}{4g_5^2z}A^{MN}A_{MN}\right] \  ,
\end{align}
where $ D X = \partial X - iA_{L}X + iX A_{R} $, ($ A_{L(R)} = V \pm A $) and $X(x,z) = v(z)U(x,z)/2 $ is taken as a
product of the chiral field $ U(x,z) =  \exp{\left(2i t^a \pi^a(x,z)\right)} $ and the function $ v(z) = m_q z +
\sigma z^3 $ containing the chiral symmetry breaking  parameters $m_q$ and $\sigma$, with  $m_q$ playing the role
of  the quark mass and $\sigma$  that of the  quark   condensate. Expanding $U(x,z) $ in powers of $ \pi^a $ gives
the relevant piece of the  action
\begin{align}
S^{A \,  (2)}_{\rm AdS} &= {\rm Tr }\int d^4x~dz~\left[-\frac{1}{4g_5^2z}A^{MN}A_{MN} + \frac{v^2(z)}{2z^3}(A^a_M
-
\partial_M\pi^a)^2 \right].
\end{align}
This Higgs-like mechanism breaks the axial-vector gauge symmetry by bringing a $z$-dependent mass term in the
$A$-part of the lagrangian.  Varying the action with respect to the transverse part of the axial-vector gauge field
$ A^a_{\perp \mu}(x,z) $ and  representing the Fourier image of $ A^a_{\perp \mu}(x,z) $ as $
\tilde{A}^a_{\perp\mu}(p,z) $ we will get the following equation of motion
\begin{align}\label{axvec}
\left[ z^3\partial_z\left(\frac{1}{z}\partial_z \tilde A^a_{\mu}\right) + p^2z^2\tilde A^a_{\mu} - g^2_5 v^2
\tilde A^a_{\mu}\right]_{\perp} = 0 \  ,
\end{align}
that determines physics of the axial-vector mesons, like $ A_1 $. The axial-vector bulk-to-boundary propagator $
{\cal A}(p,z) $ is introduced by the relation
$ \tilde{A}^a_{\perp\mu}(p,z) = {\cal A}(p,z)A^a_{\mu}(p) $. It
satisfies Eq.~(\ref{axvec}) with boundary conditions (b.c.)  $ {\cal A}(p,0) = 1 $ and $ {\cal A}'(p,z_0)= 0 $.
Similarly, variation with respect to the longitudinal component $
\partial_{\mu} \psi^a $ gives
\begin{align}\label{phieq}
z^3\partial_z\left(\frac{1}{z}\, \partial_z \psi^a \right) - g^2_5 v^2\left(\psi^a - \pi^a\right) = 0 \ .
\end{align}
Finally, varying with respect to $ A_z $  produces
\begin{align}\label{pieq}
p^2z^2 \partial_z \psi^a - g^2_5 v^2 \partial_z\pi^a = 0 \ .
\end{align}
The pion wave function is determined from Eqs.~(\ref{phieq}) and (\ref{pieq}) with b.c. $
\partial_z \psi (z_0) = 0 $, $ \psi(\epsilon) = 0 $ and $ \pi(\epsilon) = 0 $.

 Within the framework of the
model of Ref.~\cite{Erlich:2005qh},  it is possible to  derive the
Gell-Mann--Oakes--Renner relation $m_\pi^2 \sim m_q$ producing
massless pion in the $m_q=0$ limit. Taking $ p^2 =
m^2_{\pi} $ in Eq.~(\ref{pieq}) gives
\begin{align}\label{pieq1}
\partial_z\pi = \frac{m^2_{\pi}z^2 }{g^2_5 v^2}\partial_z \psi \  .
\end{align}
A perturbative solution in the form of $m_\pi^2$ expansion
was proposed in Ref~\cite{Erlich:2005qh}, with $ \psi(z) = {\cal A}(0,z) - 1 $ in the lowest order.
 Then it was  shown that, in the $m_q \to 0$ limit, $ \pi(z)$  tends to $-\theta(z - z_0) $ or,
roughly speaking, $ \pi = - 1 $ in this limit. Since our  goal is to calculate the pion form factor in the chiral
limit, this approximation will be sufficient for us.

\subsection{Two-Point Function}

The spectrum in   the  axial-current channel consists of the  pseudoscalar pion $ \langle 0|J_A^{\alpha}|\pi(p)
\rangle = if_{\pi}p^{\alpha} $ and axial-vector  mesons   $ \langle 0|J_A^{\alpha}|A_n(p, s) \rangle =
 F_{A,n}  \epsilon^{\alpha}_n (p,s)$, where $ F_{A,n} $ correspond to the $n^{\rm th} $ axial-vector meson
decay constant (and we ignored the flavor indexes). Thus, the two-point function for the axial-vector currents has
the form:
\begin{align}
\label{2axial} \langle\,  J_{A}^{\alpha}(p)J_{A}^{\beta}(-p)\, \rangle =  p^{\alpha}p^{\beta}\frac{f^2_{\pi}}{p^2}
+ \sum_n \Pi^{\alpha \beta}_n (p) \,  \frac{F^2_{A,n}}{p^2 - M^2_{A,n}}\   .
\end{align}
where the meson polarization tensor  is   given by
\begin{align}
\Pi^{\alpha \beta}_n (p) = \sum_{s} \epsilon^{\alpha}_n (p,s) \epsilon^{\beta}_n (p,s) = -
\eta^{\alpha \beta} + \frac{p^{\alpha}p^{\beta}}{M_{A,n}^2}  \  .
\end{align}
The representation for the two-point function can  be also written  as
\begin{align}
\langle \,  J_{A}^{\alpha}(p)J_{A}^{\beta}(-p)\,  \rangle =  p^{\alpha}p^{\beta}\frac{f^2_{\pi}}{p^2} + \left (-
\eta^{\alpha \beta} + \frac{p^{\alpha}p^{\beta}}{p^2}  \right ) \sum_n  \,  \frac{F^2_{A,n}}{p^2 - M^2_{A,n}}\  +
{\rm (nonpole \ terms )} ,
\end{align}
in which  the second term on the rhs is explicitly transverse to $p$.

As noted in Ref.~\cite{Erlich:2005qh}, using  holographic correspondence one can relate the two-point function to
$[\partial_z A(p,z)/z]_{z=0}$ and  derive that
\begin{align}\label{piondecay1}
f^2_{\pi} =  - \frac{1}{g^2_5}\left(\frac{1}{z}\partial_z A(0,z)\right)_{z = \epsilon \rightarrow 0}  \ .
\end{align}
For large spacelike $ p^2 $, Eq.~(\ref{axvec}) gives the same solution as in case of vector mesons, and  the same
asymptotic logarithmic behavior, just  as expected from QCD.

\subsection{Pion Wave Functions}

The longitudinal component of the axial-vector gauge field was defined as $ A_{\|} = \partial \psi $. In the
chiral limit, when $ p^2 = m_{\pi}^2 = 0 $, we have $\partial_z\pi = 0 $, and the basic equation for $\psi$,
Eq.~(\ref{phieq}) can  be rewritten as the equation
\begin{align}\label{pioneq}
z^3\partial_z\left(\frac{1}{z}\, \partial_z \Psi \right) - g^2_5 v^2 \Psi  = 0
\end{align}
for the function $ \Psi \equiv \psi - \pi $. In   the   chiral  limit, when $\pi (z) \to -1$, the value of  $
\Psi(\epsilon )$   tends to   1  as $\epsilon \to 0$. This   value  and the b.c. $ \Psi'(z_0) = 0 $ are the same
as those for ${\cal A} (p,z)$ and, furthermore, Eq.~(\ref{pioneq}) coincides with the $p^2=0$ version of equation
(\ref{axvec}) for ${\cal A} (p,z)$. Hence, the solution for $\Psi(z)$ coincides with
 ${\cal A} (0,z) $:
\begin{align}
\Psi(z) = {\cal A} (0,z) \  ,
\end{align}
and we may write
\begin{align}\label{piondecay2}
f^2_{\pi} =  - \frac{1}{g^2_5}\left(\frac{1}{z}\partial_z \Psi(z) \right)_{z = \epsilon \rightarrow 0}  \ .
\end{align}
In our analysis of $\rho$-meson wave functions in Refs.~\cite{Grigoryan:2007vg,Grigoryan:2007my}, we emphasized
that it makes sense  to  consider   also the conjugate functions $\Phi (z) \sim \Psi'(z)/z$ of the corresponding
Sturm-Liouville equation.  As we observed, they  are closer in their structure to the usual quantum mechanical
bound state wave functions than the solutions of the original equation. In the pion  case, it is  convenient to
define the  $\Phi$ function as
\begin{align}
\Phi(z) = - \frac1{g_5^2 f_\pi^2} \left ( \frac1{z} \, \partial_z \Psi (z) \right)  \  .
\end{align}
It vanishes at the IR boundary $z=z_0$ and, according to  Eq.~(\ref{piondecay1}), is  normalized as
\begin{align}
\label{psi0} \Phi(0) = 1
\end{align}
at the origin.  Note also  that  using Eq.~(\ref{pioneq})  we  can express $\Psi$ as  derivative of $\Phi$:
\begin{align}
\label{phiz} \Psi(z) = - \frac{ f_\pi^2\, z^3} {v^2} \, \partial_z \Phi (z)   \  .
\end{align}

\section{Extracting  Pion  Electromagnetic Form Factor}

\subsection{Three-point function}

To  obtain the pion  form    factor, we need to  consider three-point correlation functions. The correlator should
include the external EM current $J^{el}_\mu(0)$ and    currents having   nonzero projection onto the pion states,
e.g. the axial   currents $ J^a_{5 \alpha} (x_1) , J_{5 \beta}^{a\dagger}(x_2)$
\begin{equation} \label{threepoint}
{\cal T}_{\mu \alpha \beta }(p_1,p_2) =
 \int d^4x_1 \int d^4x_2\  e^{i p_1x_1 - ip_2 x_2} \
  \langle 0|{\cal
T}J_{5 \beta}^{\dagger}(x_2) J^{\rm el \,  }_\mu(0) J_{5\alpha} (x_1)|0\rangle \ ,
\end{equation}
where $p_1 ,  p_2 $ are  the   corresponding momenta, with  the momentum transfer carried by the EM source being $
q = p_2 - p_1 $ (as usual, we denote $ q^2 = - Q^2 $,  $ Q^2 > 0 $). The spectral representation for the
three-point  function is a  two-dimensional generalization of Eq.~(\ref{2axial})
\begin{align}
{\cal T}^{\mu \alpha \beta }(p_1,p_2) = p_1^{\alpha}p_2^{\beta} (p_1+p_2)^\mu
 \frac{f^2_{\pi}\, F_{\pi}(Q^2)}{p_1^2 p_2^2 } +
\sum_{n,m}  {\rm (transverse \ terms) } \ + {\rm (nonpole \ terms) }   ,
\end{align}
where the first term, longitudinal both with respect to $p_1^\alpha$ and $p_2^\beta$ contains the pion
electromagnetic form factor $F_{\pi}(Q^2)$
\begin{equation}
\langle\pi(p_1)|J^{el}_\mu(0)|\pi(p_2)\rangle = F_\pi(q^2)(p_1+p_2)_\mu \ ,
\end{equation}
(normalized by $ F_\pi(0)=1 $), while other pole terms contain the contributions involving axial-vector mesons and
are transverse either with   respect to $p_1^\alpha$ or $p_2^\beta$, or  both. Hence,    the pion form   factor
can be extracted from the three-point   function using
\begin{align}
\label{proj0} p_{1\alpha} p_{2\beta} {\cal T}^{\mu \alpha \beta }(p_1,p_2) |_{p_1^2=0, p_2^2 =0}  =
 (p_1+p_2)^\mu
{f^2_{\pi}\, F_{\pi}(Q^2)} \   .
\end{align}

\subsection{Trilinear Terms in $F^2$ Part of  Action}

To obtain  form factor from the holographic model, we need the action of the third order in the fields. There are
two types  of terms   contributing to the pion electromagnetic  form factor: $|DX|^2$ term  and $F^2$  terms. Let
us   consider first the contribution  from $F^2$ terms.  They contain $ VVV $, $ VAA $ and $ AVA $ interactions
and  may   be written as
\begin{align}
S^{F^2 }_{\rm AdS} |_3 = \frac{i}{g_5^2}\, {\rm Tr} \int d^4x~dz~\frac{1}{z}\left(V_{MN}[V^{M},V^{N}] +
V_{MN}[A^{M},A^{N}] +A_{MN}[V^{M},A^{N}] \right)
 \ ,
\end{align}
where $ V_{MN} =
\partial_{M}V_{N} -
\partial_{N}V_{M} $ and $ A_{MN} =
\partial_{M}A_{N} -
\partial_{N}A_{M} $.  Taking  $ V_{z} =
A_{z} = 0 $ gauge, we pick out the part of the action which is 
contributing to the 3-point function $ \langle
J_{5\alpha}J_{\mu}J_{5\beta} \rangle $:
\begin{align}
\label{xnia}
W_{3} = \frac{i}{g_5^2}\, {\rm Tr} \, \int d^4x~dz~\frac{1}{z}\left(V_{\mu\nu}[A^{\mu},A^{\nu}] +
A_{\mu\nu}[V^{\mu},A^{\nu}] \right).
\end{align}
Introducing  Fourier transforms of fields,  we  define, as usual,   $ V_{\mu}(q,z) = \tilde{V}_{\mu}(q){\cal V}
(q,z) $ for the  vector field, where  $\tilde{V}_{\mu}(q)$  is the Fourier transform of the 4-dimensional field
$V_{\mu} (x)$ and  ${\cal V} (q,z)$ is the bulk-to-boundary propagator satisfying the equation
\begin{equation}
z \,  \partial_z\left(\frac{1}{z}\, \partial_z {\cal V}(q,z)  \right)
 + q^2\,  {\cal V} (q,z) =0
\end{equation}
with b.c. ${\cal V}(q,0) = 1 $ and $ \partial_z{\cal V}(q,z_0) = 0 $. It can be written as the sum
\begin{equation}
\label{Jmeson}  {\cal V } (q,z) = g_5\sum_{m = 1}^{\infty}\frac{ f_{m} \psi_m^V( z)}{ -q^2 + M^2_{m} }
\end{equation}
involving  all the  bound states in the $q$-channel, with $M_m$ being the mass of the $m$th bound state and  $
\psi_m^V(z) $ its wave  function given by a solution of the basic  equation of   motion in the  vector sector.

The projection (\ref{proj0}) picks out only the longitudinal part $A_{\parallel \mu} (p,z)$ of  the axial-vector
field. Taking  into account that $ A^a_{\parallel \mu}(x,z) = \partial_{\mu}\psi(x,z) $, we   may  write
\begin{align}
A^a_{\parallel \mu}(p,z) =  ip_{\mu}\psi^a(p,z) \ ,
\end{align}
where $A^a_{\parallel \mu}(p,z)$ and $\psi^a(p,z)$  are the Fourier transforms of 
$ A^a_{\parallel \mu}(x,z)$ and $\psi(x,z) $, respectively.
Furthermore, there is only one particle in the  expansion over  bound state in this   case,   namely,  the massless
pion. Thus, we have $A^a_{\parallel \mu}(p,z) = \tilde{A}^a_{\parallel \mu}(p) \, \psi (z)$ and, therefore,
\begin{align}
\psi^a(p,z) = -\frac{ip^{\alpha}}{p^2}\tilde{A}^a_{\parallel \alpha}(p)\psi(z) \  .
\end{align}
This allows us to rewrite $ A^a_{\parallel \mu}(p,z) $ in  the form
\begin{align}
A^a_{\parallel \mu}(p,z) = \frac{p^{\alpha}p_{\mu}}{p^2}\tilde{A}^a_{\parallel \alpha}(p)\, \psi(z)
\end{align}
involving the longitudinal projector $p^{\alpha}p_{\mu}/{p^2}$  and the pion wave function $\psi(z)$, which is the
solution of the basic equation (\ref{phieq}). Using this representation and making Fourier transform of $ W_3 $
gives
\begin{align}
W_{3} &= -\frac{1}{2g_5^2} \, \epsilon_{abc} \, \int \frac{d^4u \, d^4v \,  d^4w}{(2\pi)^{12}} \ i(2\pi)^4
\delta^{(4)}(u+v+w)\frac{u^{\mu}v^{\nu}u^{\alpha}v^{\beta}}{u^2 v^2} \\
\nonumber & \tilde{A}^b_{\parallel \alpha}(u)\tilde{A}^c_{\parallel \beta}(v)\left[w_{\mu}\tilde{V}^a_{\nu}(w) -
w_{\nu}\tilde{V}^a_{\mu}(w) \right] \int^{z_0}_{\epsilon} dz~\frac{1}{z}\, {\cal V} (w,z)\, \psi^2(z) 
\end{align}
(notice that the second term in Eq.(\ref{xnia}) vanishes for longitudinal
axial-vector fields).
Varying this functional with respect to sources produces  the following 3-point function:
\begin{align}
\langle J_{V,a}^{\mu}(q)J_{\parallel A,b}^{\alpha}(p_1) J_{\parallel A,c}^{\beta}(-p_2) \rangle &=
 -i(2\pi)^4\delta^{(4)}(q + p_1 - p_2) \,\epsilon_{abc} \ \frac{p_1^{\alpha}p_2^{\beta}}{p^2_1p^2_2}\,
(p_1+p_2)^{\mu}
\nonumber \\
&  \times  \frac{1}{2g^2_5}\, q^2 \int^{z_0}_{\epsilon} dz~\frac{1}{z}\, {\cal V} (q,z)\, \psi^2(z) \ ,
\end{align}
where, anticipating the limit $p^2_1 \rightarrow 0, \, p^2_2 \rightarrow 0$, we took  $ (p_1q) = -(p_2q) = -q^2/2
$ in the numerator  factors. Now,  representing  $ \langle J_{V,a}^{\mu}(q) J_{\parallel A,b}^{\alpha}(p_1)
J_{\parallel A,c}^{\beta}(-p_2) \rangle = i(2\pi)^4\delta^{(4)}(q + p_1 - p_2)\, \epsilon_{abc}{\cal T}^{\mu
\alpha \beta }(p_1,p_2)\  $ and applying the projection suggested by Eq.~(\ref{proj0}), we will have
\begin{align}\label{proj}
& \lim_{p^2_1 \rightarrow 0} \ \lim_{p^2_2 \rightarrow 0} p_{1\alpha}p_{2\beta}{\cal T}^{\mu \alpha \beta
}(p_1,p_2) = \frac{1}{2g^2_5} \ (p_1+p_2)^{\mu} Q^2 J(Q) \ ,
\end{align}
where $J(Q)$ is the dynamic factor   given by the convolution
\begin{align}
&J(Q) = \int^{z_0}_{\epsilon} {\frac{dz}{z}} \, {\cal J} (Q,z) \, \psi^2(z) \ .
\end{align}

\subsection{Dynamic Factor and Wave Functions}

The vector   bulk-to-boundary propagator $ {\cal J}(Q,z) \equiv {\cal V } (iQ,z)$ for spacelike momenta, entering
into the dynamic factor $J(Q)$, satisfies the equation
\begin{equation}
z \,  \partial_z\left(\frac{1}{z}\, \partial_z {\cal J}(Q,z)  \right)
 = Q^2\,  {\cal J} (Q,z)
\end{equation}
with b.c. ${\cal J}(Q,0) = 1 $ and $ \partial_z{\cal J}(Q,z_0) = 0 $.  Its explicit form  is given  by
\begin{equation}
{\cal J} (Q,z) = {Qz}\left[K_1(Qz) + I_1(Qz) \frac{K_0(Qz_0)}{I_0(Qz_0)} \right] \ .
\end{equation}
One can easily see  that ${\cal J}(0,z) = 1$. Combining all the factors, we get
\begin{align}
f^2_{\pi} F_\pi^{(F^2)} (Q^2)   &= \frac{1}{2g^2_5}\, Q^2 \int_{0}^{z_0} \frac{dz}{z}\, {\cal J} (Q,z)\, \psi^2(z)
\ .
\end{align}
Integrating  by parts  and using  equations of motion  both for $ {\cal J} $ and $ \psi $ gives
\begin{align}
\label{almost} F_\pi^{(F^2)} (Q^2) &=  \frac{1}{g^2_5f^2_{\pi}} \int_{0}^{z_0} {dz}\, {z}\, {\cal J}
(Q,z)\,\left[\left(\frac{\partial_z\psi}{z}\right)^2 + \frac{g^2_5 v^2}{z^4}\, \psi \,  (\psi - \pi) \right]    \
.
\end{align}
We need  also to add the  $V\pi \pi$ contribution from the $|DX|^2$ term of  the AdS action (\ref{AdS}). It is
generated by
\begin{align}
S^{|DX|^2}_{\rm AdS}  |_ {V\pi \pi}&= \epsilon_{abc} \int d^4x~dz~\left[ \frac{v^2(z)}{z^3}(A^a_M -
\partial_M\pi^a)\,  \pi^b \,  V^{c\, M} \right],
\end{align}
and its inclusion changes $\psi (\psi - \pi)$ into $(\psi - \pi)^2$ in Eq.~(\ref{almost}). 
The total result   (see also Ref.~\cite{Erlich:2005qh})  may
be now conveniently expressed
 in terms of the $\Psi \equiv  \psi - \pi$
wave function
\begin{align}
\label{FF}
 F_\pi(Q^2) &= \frac{1}{g^2_5 f_\pi^2}\int_{0}^{z_0} dz
  \, z \, {\cal J} (Q,z)\left[\left(\frac{\partial_z\Psi}{z}\right)^2 +
\frac{g^2_5 v^2}{z^4} \Psi^2(z)\right] \ .
\end{align}
Using  equation of motion for $ \Psi(z) $,  one can see that the expression   in square brackets coincides with
$$\frac1z \, \partial_z \left (\Psi (z) \, \frac1z \, \partial_z\Psi (z) \right )=  - g_5^2 f_\pi^2 \, \frac1z \,
\partial_z \Bigl  (\Psi(z)  \, \Phi  (z) \Bigr  )  \   , $$
 and write the form  factor as
\begin{align}
\label{FF3}
 F_\pi(Q^2) &= - \int_{0}^{z_0} dz
  \ {\cal J} (Q,z) \, \partial_z \Bigl  (\Psi(z)  \, \Phi  (z)
\Bigr  )
   \ .
\end{align}
  This  representation allows one to easily check the
 normalization
\begin{align}\label{norm}
F_\pi(0) = - \int_{0}^{z_0} dz \,
 \partial_z  \Bigl  (\Psi(z)  \, \Phi  (z)
\Bigr  )  =  \Psi(0) \, \Phi (0)   = 1 \ ,
\end{align}
where we took into account that $ {\cal J } (0,z) = 1 $ and $\Phi (z_0) =0$. We can also represent our result for
the pion form factor as
\begin{align}
\label{FF2}
 F_\pi(Q^2) &=\int_{0}^{z_0} dz
  \, z \, {\cal J} (Q,z) \left[{g^2_5 f_\pi^2} \Phi^2 (z)  + \frac{\sigma^2}
{f^2_{\pi}} \, z^2\, \Psi^2(z) \right] \equiv \int_{0}^{z_0} dz
  \ z \ {\cal J} (Q,z)\, \rho (z) \ ,
\end{align}
and interpret the function $\rho(z) $
 as the  radial distribution density, as it was done in
Refs.~\cite{Grigoryan:2007vg,Grigoryan:2007my}. Note that keeping only the   first term  in square brackets gives
an
 expression similar to our result  \cite{Grigoryan:2007vg} for the $\rho$-meson
form factor
\begin{align}
\label{calf}
 {\cal F}_{11}(Q^2)= \int_0^{z_0} dz \, z  \, {\cal J} (Q,z) \, |\phi_1(z)|^2
\end{align}
in terms of the function $\phi_1$   conjugate to the solution of the basic equation of motion.   The  value of
$\phi_1 (z)$ at the origin is proportional to  the $\rho$-meson decay constant $f_\rho/m_\rho \equiv g_\rho$
(experimentally, $g_\rho^{\rm exp} \approx 207 \, {\rm MeV}$ \cite{Eidelman:2004wy}), 
namely,  $\phi_1 (0) =g_5\, g_\rho$. Thus, the pion
wave function $g_5 f_\pi \Phi (z)\equiv \phi_\pi (z)$ is  a direct   analog of the $\rho$-meson wave  function
 $\phi_1(z)$.
Main difference is that,  in the pion   case,   there  is also the second term in the form factor expression. The
latter, in fact,  is  necessary to secure correct normalization of the form  factor at  $Q^2=0$.   In
Eq.~(\ref{FF}),   this   term is written  in   terms of the $\Psi (z) $  wave   function,
 but using Eq.~(\ref{phiz}) we can rewrite it
also in terms of $\Phi (z)$ or $\phi_\pi (z)$:
\begin{align}
\label{rhopsiz} \rho(z)  =\phi _\pi^2 \,
 (z)  + \frac1{g_5^2\sigma^2} \,  \left ( \frac1{z^2}\,  \partial_z \phi_\pi (z) \right )^2 \  .
\end{align}

\section{Wave Functions and Form Factor}

 \subsection{Structure of Pion Wave Functions}

Explicit form of the $\Psi$   wave function  follows from  the
solution of Eq.~(\ref{pioneq}):
\begin{align}
\Psi (z) =
{z\, \Gamma \left [{2}/{3} \right  ]
\left(\frac{\alpha}{2}\right)^{1/3}}
\left[ I_{-1/3}\left(\alpha z^3\right)  -  I_{1/3}\left(\alpha
z^3\right) \frac{I_{2/3}\left(\alpha z^3_0\right)} {I_{-2/3}\left(\alpha z^3_0\right)}\right]  \ ,
\end{align}
where $ \alpha = g_5 \sigma/3 \approx 1.481 \, \sigma $
(recall that $g_5=\sqrt{2}  \pi$, see e.g. Ref.\cite{Grigoryan:2007my}).
As a result,  $\Phi(z)$  is given by
\begin{align}
\label{Psi}
\Phi(z) = - \frac1{g_5^2 f_\pi^2} \left ( \frac1{z} \,
\partial_z \Psi  (z) \right)  \   =
\frac{3\, z^2}{g_5^2 f_\pi^2 }\, \Gamma\left[2/3\right]
\left(\frac{\alpha^4}{2}\right)^{1/3}
\left[- I_{2/3}\left(\alpha
z^3\right)  +  I_{-2/3}\left(\alpha
z^3\right)
\frac{I_{2/3}\left(\alpha z^3_0\right)} {I_{-2/3}\left(\alpha
z^3_0\right)}\right]  \ .
\end{align}
This  formula, combined with  Eq.~(\ref{psi0}),
 establishes the relation
\begin{align}
\label{fpi2}
 f_\pi^2 = 3\cdot 2^{1/3} \, \frac{\Gamma[2/3]}{\Gamma[1/3]}\,
\frac{I_{2/3}\left(\alpha z^3_0\right)} {I_{-2/3}\left(\alpha
z^3_0\right)}   \frac{\alpha^{2/3}}{g_5^2} 
\end{align}
for  $f_\pi$  in terms of the  condensate
parameter $\alpha$ and the
confinement radius $z_0$.   Since $\sigma$  appears in the solutions
only through $\alpha$, we will use  $\alpha$ in what follows. Note also that
$\alpha^{1/3} \approx 1.14\, \sigma^{1/3}$.

\begin{figure}[h]
\mbox{
  {\epsfysize=4cm  \epsffile{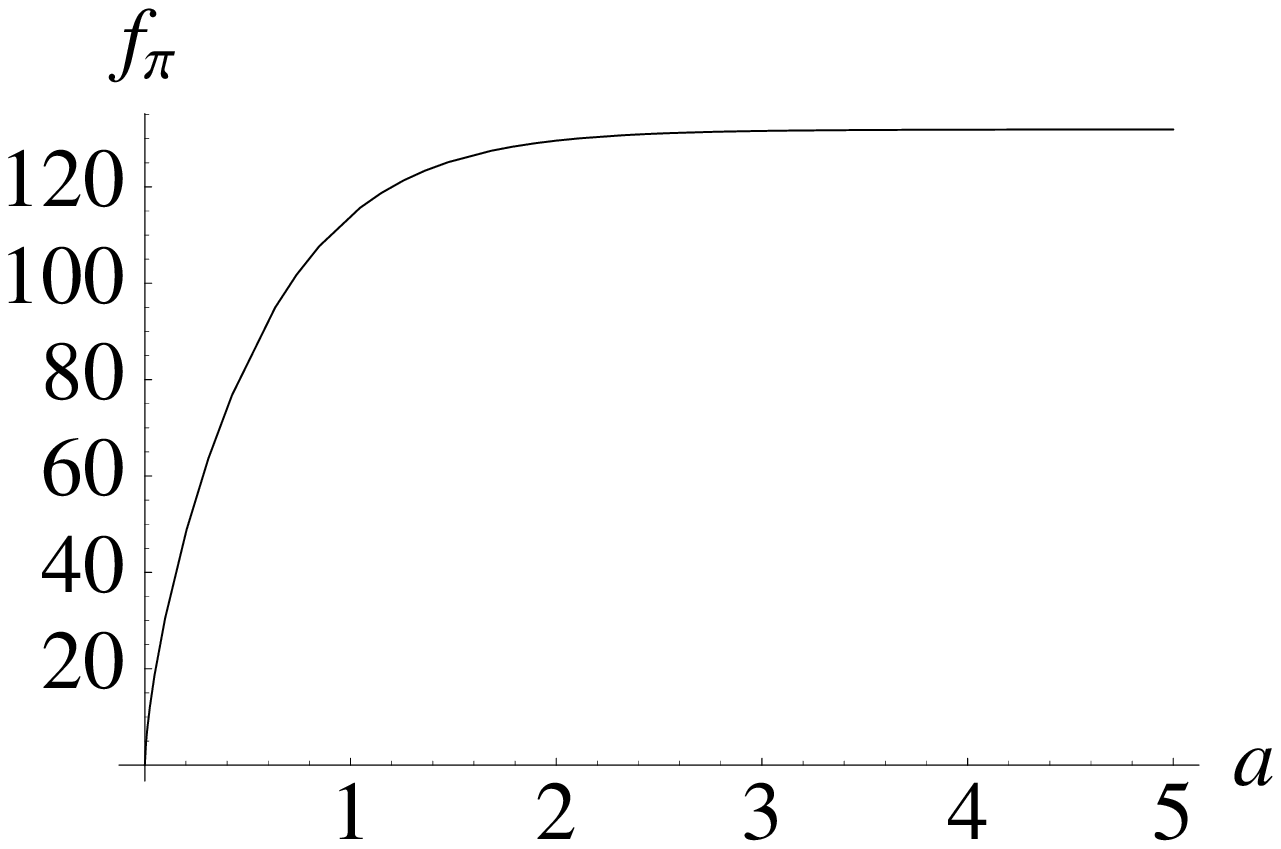}} \hspace{1cm}
{\epsfysize=4cm  \epsffile{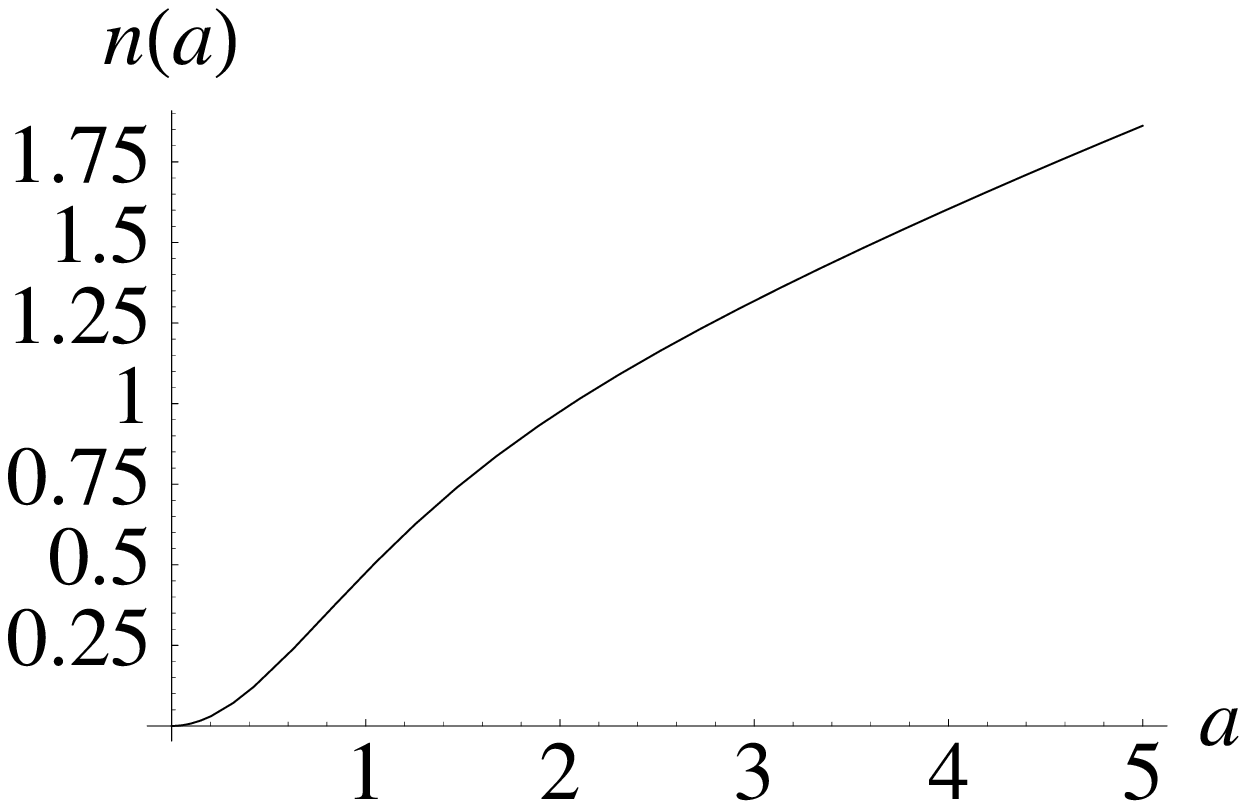}} }
\caption{\label{fpi}
{\it Left:}  Pion decay constant $f_\pi$ as a function of $a$
for fixed $\alpha^{1/3}=424\,$MeV. {\it Right:} Function $n(a)$
}
\end{figure}

Realizing that the equations of motion for  the vector sector in this  holographic model
are  not affected by  the   chiral symmetry-breaking effects expressed through the
function $v(z)$, it is
natural to  set  the value of $z_0$
from the vector sector spectrum,
i.e.,  by the $\rho$-meson mass. The   numerical   value of
  $z_0$ (call it $z_0^\rho$)  is then
$z_0^\rho  \approx 1/323 \, {\rm MeV}$.
As given by Eq.~(\ref{fpi2}), $f_\pi$  looks
like a rather complicated function
of two scales, $z_0$ and $\alpha$. Note, however,
that   the  ratio $I_{2/3}(a) /  I_{-2/3} (a)$ is very   close to 1 for $a \gtrsim 2$ and
practically indistinguishable
from 1 for $a \gtrsim 3$.  Hence,   for sufficiently  large values of the
confinement radius,  $z_0 \gtrsim 1/ \alpha^{1/3}$,
the value of $f_\pi$ is    determined
by the value of $\alpha$  alone.
This  limiting   value of $f_\pi$ is given by
\begin{align}
\label{fpiinfty}
  f_\pi  |_{z_0 \to \infty} =2^{1/6} \, \frac{ \alpha^{1/3}}{g_5} \,
 \sqrt{ \frac{3 \Gamma[2/3]}{\Gamma[1/3]}}
= \frac{3^{1/2}}{2^{1/3} \pi} \,
\sqrt{ \frac{ \Gamma[2/3]}{\Gamma[1/3]}} \,  \alpha^{1/3} \,
\approx \frac{\alpha^{1/3}}{3.21}  \ .
\end{align}
Requiring that $f_\pi  |_{z_0 \to \infty}$
coincides with the experimental value, $f_\pi \approx 131\, {\rm MeV}$,
one should take  $\alpha^{1/3} \approx 420 \, {\rm MeV}$.
For such  $\alpha$, the value of $ 1/ \alpha^{1/3}$  is close to    $z_0^\rho$,
i.e., we are in the region $\alpha z_0^3  \sim 1$
and we may expect that, even if we use
exact  formula  (\ref{fpi2}) with $z_0=z_0^\rho$,  the value of $f_\pi$
would not change much.  Indeed, to get
$f_\pi \approx 131\, {\rm MeV}$
from  Eq.~(\ref{fpi2})
for $1/z_0 = 323 \, {\rm MeV}$, we should take
 $\alpha^{1/3} \approx 424 \, {\rm MeV}\equiv \alpha_0^{1/3}$.
Thus, in this range of parameters,  the  value of $f_\pi$ is practically  in one-to-one correspondence
with the value of $\alpha$.  It is  convenient to introduce a dimensionless
variable
\begin{align}
 a \equiv \alpha z_0^3 = \frac13 \, g_5 \sigma  z_0^3 \  .
\end{align}
Then the  values $\alpha_0^{1/3} = 424\, {\rm MeV}$ and
$1/z_0^\rho  = 323\,{\rm MeV}$  correspond to $a=2.26\equiv a_0$.
As one can see   from Fig.(\ref{fpi}),  the dependence of $f_\pi$ is practically flat
for  $a \gtrsim 2$.

The confinement   radius $z_0$ presents a  natural
scale to   measure length, so  it makes sense  to rewrite the form
factor formula (\ref{FF}) as an integral
over the dimensionless variable $\zeta \equiv z/z_0$:
\begin{align}
\label{FFxi}
 F_\pi(Q^2) &= 3 \int_{0}^{1} d\zeta
  \,  \zeta  \ {\cal J} (Q,\zeta , z_0)\left[ n(a)\, \varphi^2(\zeta ,a) +
\frac{a^2 \zeta ^2}{n(a)} \,  \psi^2(\zeta ,a)\right]  \equiv  \int_{0}^{1} d\zeta
  \,  \zeta  \ {\cal J} (Q,\zeta, z_0) \, \rho (\zeta ,a) \ ,
\end{align}
where  the mass scale   $\alpha$ is  reflected by the  dimensionless parameter $a$.
 The
   factor  $n(a)$   takes care of the  correct normalization of the  form factor.
It is  given by
\begin{align}
 n(a) = 2^{1/3} \, a^{2/3} \, \frac{\Gamma[2/3]}{\Gamma[1/3]}
 \, \frac{I_{2/3} (a)}{I_{-2/3}(a)}  \  .
\end{align}
For   small $a$,   it  may  be approximated by $\frac34 a^2$.
For  large $a$,   using  the fact that
 $I_{2/3}(a) /  I_{-2/3} (a) $ is very   close to 1 for $a \gtrsim 2$,
we may approximate  $n(a) \approx 0.637\, a^{2/3}$ in this region.
In terms of $n(a)$, the pion decay   constant  can  be written  as
\begin{align}
 f_\pi = \frac1{\pi a^{1/3}} \, \sqrt{\frac32 \,
  n(a)} \,  \alpha^{1/3}
\,    .
\end{align}
For large $a$, this gives
\begin{align}
\label{fpilarge}
 f_\pi
\,  \bigl |_{a \gtrsim 2} \approx  0.311\,  \alpha^{1/3}  \, .
\end{align}
For small $a$, we have
\begin{align}
 f_\pi \biggl  |_{a \lesssim 1}= \frac{3\,a^{2/3}}{2\sqrt{2}\pi}\,\alpha^{1/3} + \ldots
 \approx  0.338 \,  \alpha z_0^2 =  0.338 \,  \frac{a}{  z_0} \  .
\end{align}

\begin{figure}[h]
\mbox{
 {\epsfysize=4cm
 \hspace{0cm}
  \epsffile{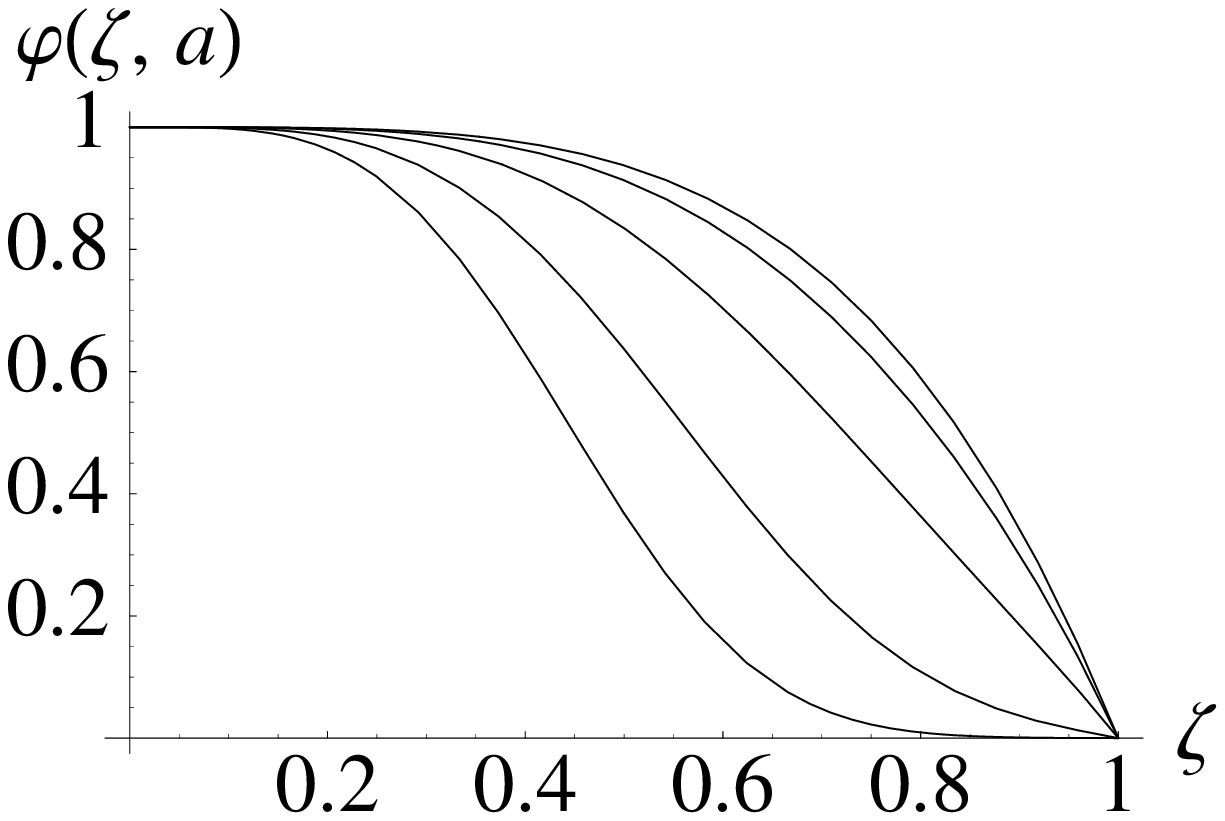} }
  \hspace{1cm}
  {\epsfysize=4cm  \epsffile{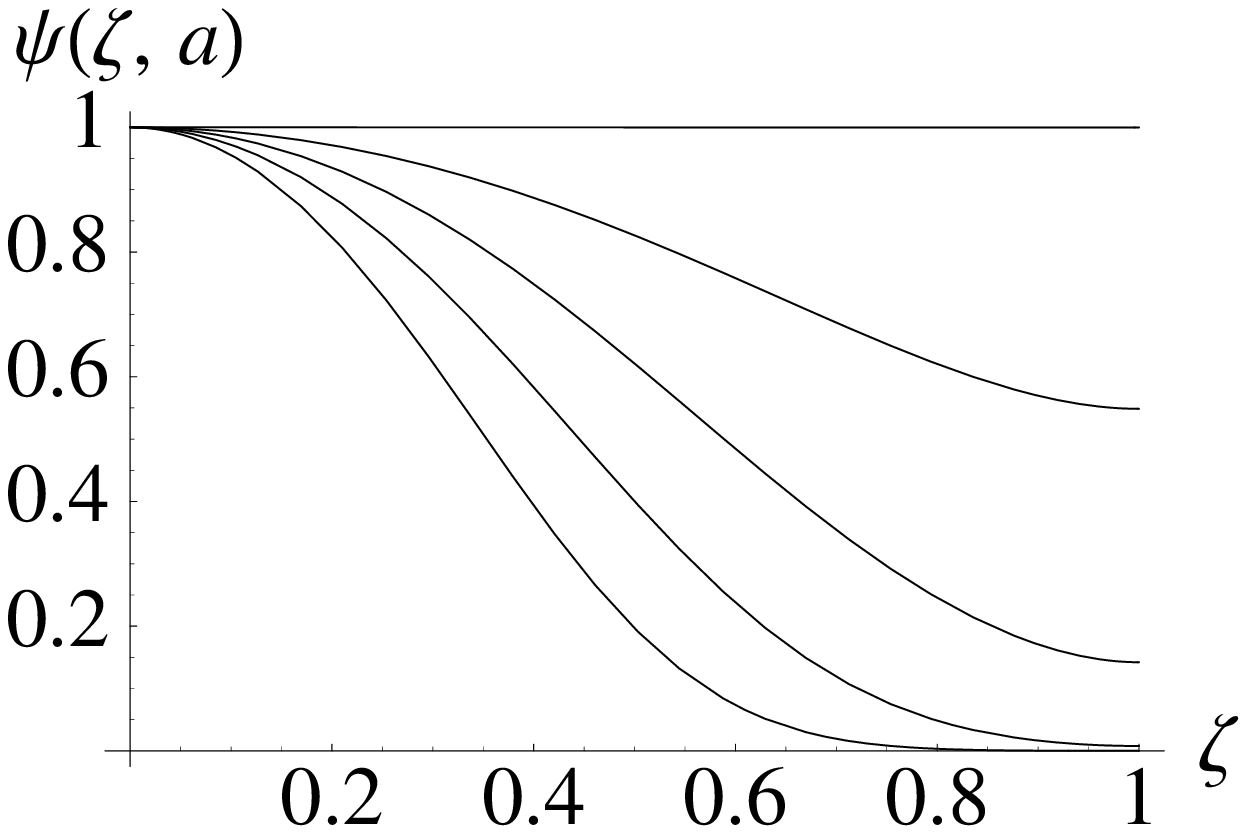}} }\hspace{0cm}
\caption{\label{psiphi}
Functions  $\varphi(\zeta ,a)$ (left) and  $\psi(\zeta ,a)$ (right)
 for several values of $a$:
$a=0$ (uppermost lines), $a=1$, $a=2.26$, $a=5$, $a=10$ (lowermost lines).
}
\end{figure}

\begin{figure}[h]
\mbox{
  {\epsfysize=3.5cm  \epsffile{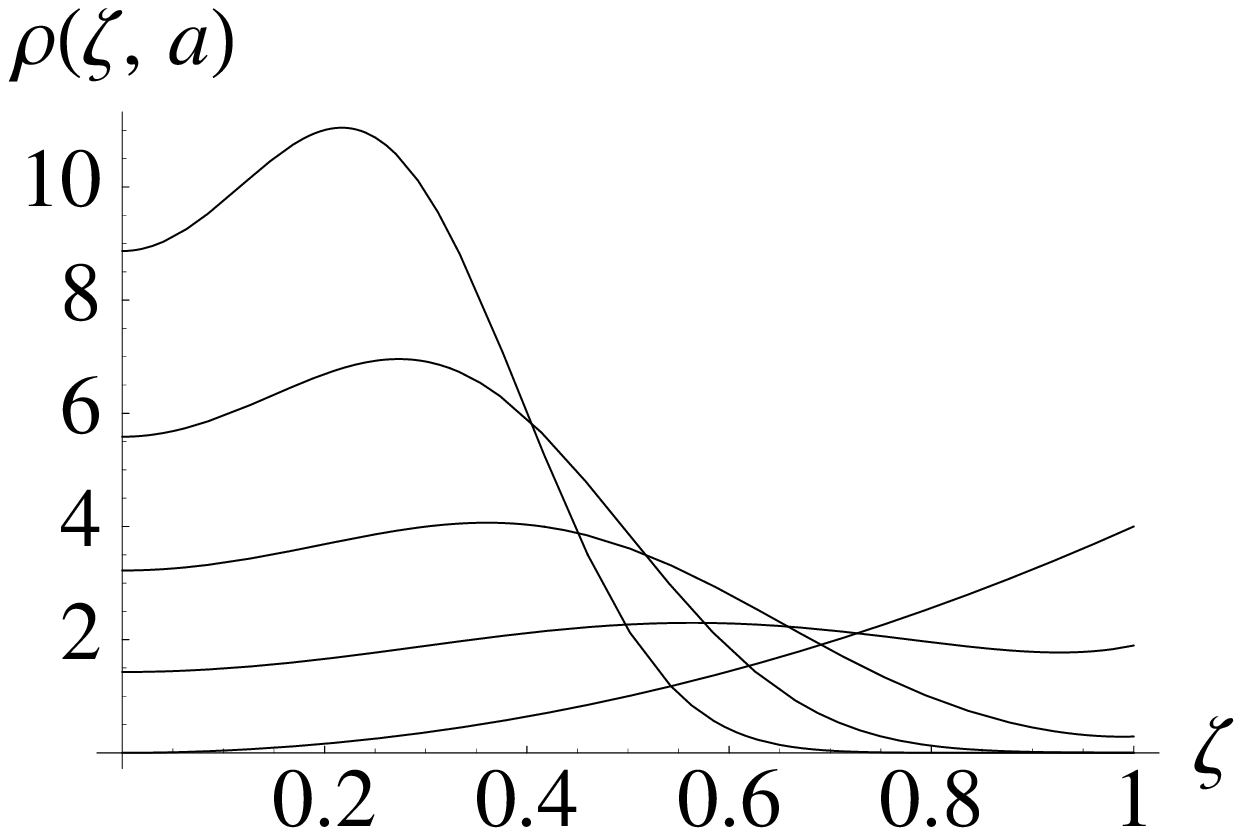}} 
{\epsfysize=3.5cm  \epsffile{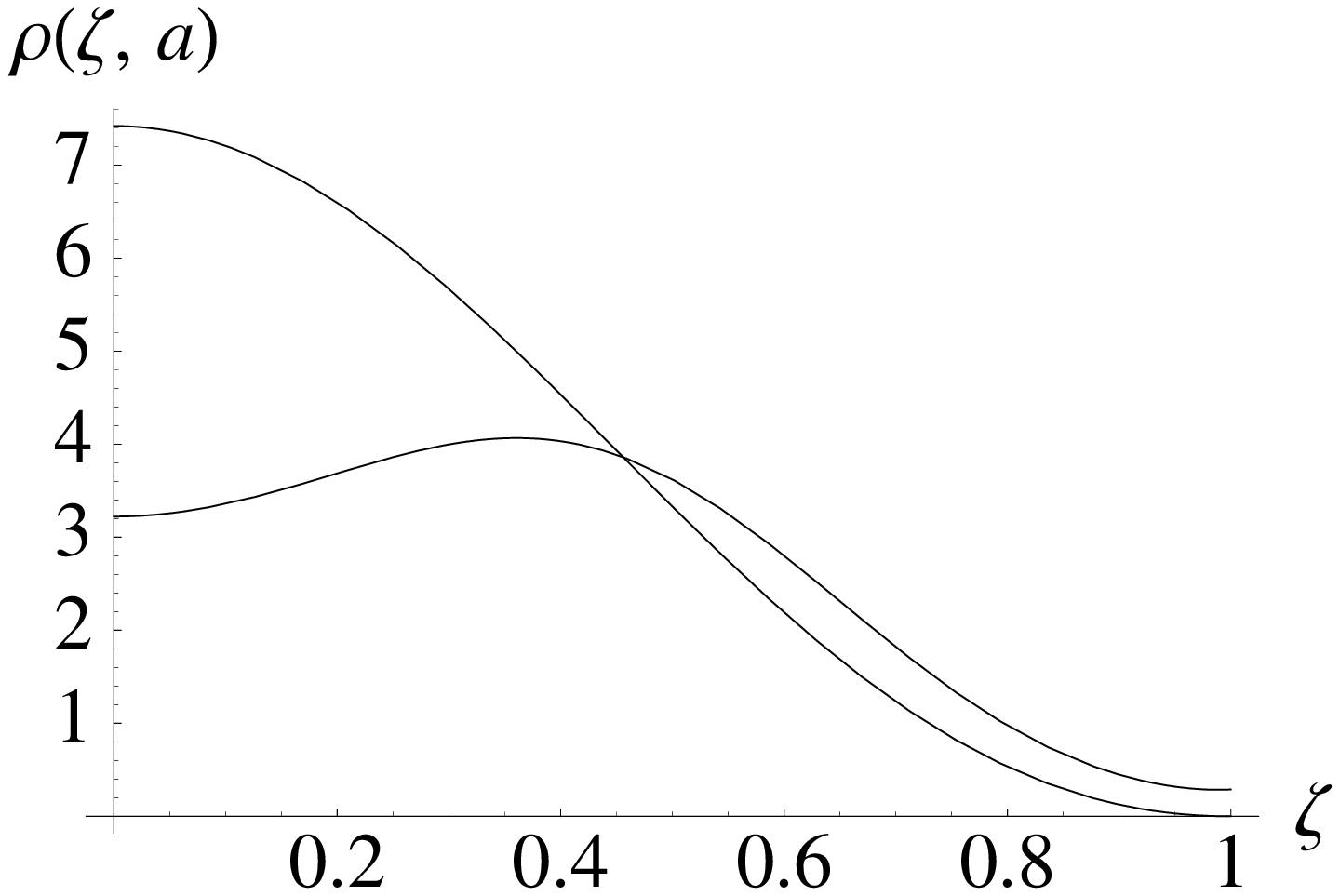}}
{\epsfysize=3.5cm  \epsffile{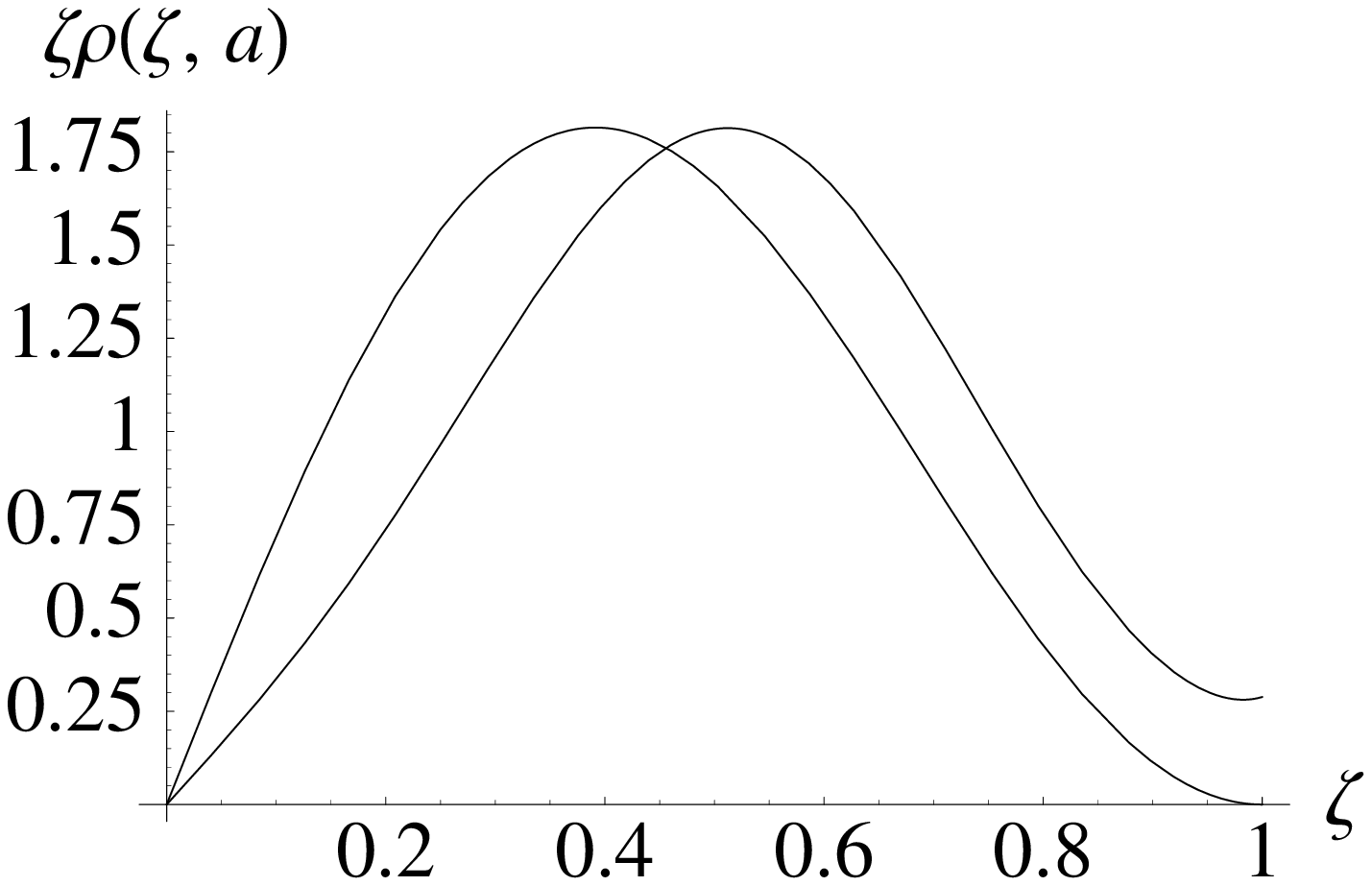}}
  }
\caption{\label{rhodense}
{\it Left:}  Function $\rho (\zeta ,a)$
 for
$a=0$, $a=1$, $a=2.26$, $a=5$, $a=10$.
{\it Middle:} Densities $\rho (\zeta ,2.26)$ for pion and $\rho_\rho  (\zeta )$
for  $\rho$-meson in the hard-wall
model.  {\it Right:} Same for densities multiplied by $\zeta$.
}
\end{figure}

 The functions $ \varphi(\zeta ,a), \psi(\zeta ,a)$ are just the $\Phi$ and $\Psi$ wave
functions written in $\zeta $  and $a$ variables.
For $a=0$, the limiting forms are $\varphi(\zeta ,0)= 1-\zeta^4$ and
$\psi (\zeta ,0)= 1$. As $a$ increases, both functions
become  more and more narrow (see Fig.\ref{psiphi}).
For density, we have $\rho (\zeta, a=0)=4 \zeta^2$ in the $a \to 0$ limit, a function  that
vanishes at the  origin (see Fig.(\ref{rhodense})). For nonzero $a$,  the value of $\rho (\zeta=0,a)$
monotonically  increases with $a$, and the function  itself narrows.
The increase of $\rho (\zeta=0,a)$ with $a$ is  generated by  the monotonically increasing
  function $n(a)$.   It is  interesting to compare the pion density $\rho (\zeta, 2.26)$
(taken at the ``experimental'' value $a=2.26$)
with the $\rho$-meson density $\rho_\rho  (\zeta )$  of Ref.~\cite{Grigoryan:2007vg}.
These densities  are rather close for $\zeta >0.5$,  but strongly differ
for small $\zeta$. In particular,  the $\rho$-meson density is
more than two times larger for $\zeta=0$,  which corresponds
 to the hard-wall model result that
$g_\rho$ is essentially larger than $f_\pi$.

\subsection{Pion Charge Radius}

It is interesting to investigate how well  these values  $z_0=1/323\, {\rm MeV}$ and
$\alpha = (424\, {\rm MeV})^3$
describe another important low-energy  characteristics of the pion --
its charge radius.
Using the $Q^2$-expansion of  the vector source \cite{Grigoryan:2007vg}
\begin{align}
 {\cal J} (Q, \zeta,  z_0)= 1 - \left. \left.  \frac{Q^2}{4}  \, z_0^2 \,
 \zeta ^2  \, \right [1- 2 \, \ln \, \zeta \right ]
+ \ldots
\end{align}
and explicit form of the density
\begin{align}
\label{rhozeta}
 \rho (\zeta, a) &= \frac32\, \Gamma (1/3)\, \Gamma (2/3)\,{a^2 \zeta^4}
\, \left [ \left ( \nu (a) \,  I_{-2/3} (a \zeta^3) -
\frac{I_{2/3} (a \zeta^3)}{\nu(a)}  \right )^2
+  \left (\frac{ I_{-1/3} (a \zeta^3)}{\nu(a)} - {\nu(a)} I_{1/3} (a \zeta^3) \right )^2
\right ] \ ,
\end{align}
where $\nu (a) \equiv \sqrt{ I_{2/3} (a)/I_{-2/3} (a)}$,
we obtain for  the pion charge radius:
\begin{align}
\label{rpi}
 \langle r^2_{\pi} \rangle = \frac32 \, z_0^2  \left. \left. \int_{0}^{1} d\zeta
  \,  \zeta ^3\,  \right [1- 2 \, \ln \, \zeta \right ]  \, \rho (\zeta ,a)
  = \frac43 \,  z_0^2 \, \left \{ 1 - \frac{a^2}{4} + {\cal O} (a^4)  \right \}  \ .
\end{align}
\begin{figure}[h]
\mbox{
  {\epsfysize=4cm  \epsffile{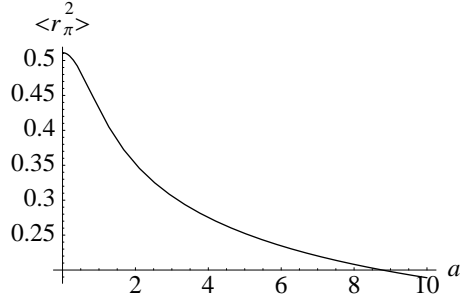}} }\hspace{0cm}
\caption{\label{radius}
$\langle r^2_{\pi} \rangle$ in fm$^2$
for $z_0 = z_0^\rho$   as a function of $a$.
}
\end{figure}

Hence, for fixed $z_0$ and small $a$,  when $\alpha  \ll 1/z_0^3$,  the pion
radius is basically determined
by the confinement scale $z_0$.  In particular,
$\langle r^2_{\pi} \rangle =\frac43 \, z_0^2$
for $\alpha =0$.
Numerically, taking   \mbox{$z_0 = z_0^\rho \approx 1/323\,{\rm MeV}=0.619\,{\rm fm}$,}
we obtain
 $\langle r^2_{\pi} \rangle = 0.51\, {\rm fm}^2 $.
This result is  very close to  the
value $\langle r^2_{\rho} \rangle_C \approx 0.53\, {\rm fm}^2 $ that we obtained in the hard-wall model for the $\rho$-meson
electric radius   determined in \cite{Grigoryan:2007vg} from the slope of the $G_C(Q^2)$ form
factor.  However, since  $G_C(Q^2)$ involves kinematic-type terms $Q^2/m_\rho^2$,
it    seems more  appropriate to compare
$F_\pi (Q^2)$ with the ${\cal F}_{11} (Q^2)$ form factor (\ref{calf})
given  directly by a wave function overlap  integral.  The slope  of  ${\cal F}_{11} (Q^2)$
is  smaller than that of $G_C(Q^2)$, and the corresponding radius is also smaller:
 $\langle r^2_{\rho} \rangle_{\cal F}  = 0.27\, {\rm fm}^2 $.
Thus, for $\alpha=0$,  the pion r.m.s. radius  is  about 1.4 times larger
 than  the $\rho$-meson size determined
by $\langle r^2_{\rho} \rangle_{\cal F}^{1/2}$.

With the increase of $\alpha$,   the pion  becomes smaller (see Fig.\ref{radius}).
The experimental value of $0.45\, {\rm fm}^2 $
\cite{Eidelman:2004wy}
 is reached  for
$a \sim 0.9$. However, the corresponding value
$f_\pi \approx 80\,{\rm MeV}$  is too small.
If  we take $a=a_0=2.26$, then $\langle r^2_{\pi} \rangle = 0.34\, {\rm fm}^2 $.
Thus, if we insist on using $z_0=z_0^\rho$ dictated by the hard-wall model calculation
of the $\rho$-meson mass, and the value   of  $\alpha$ producing the experimental
$f_\pi$ (note that then $\alpha^{-2/3} \approx 0.222\, {\rm fm}^2$),
the pion radius is smaller than the experimental value.
In  linear units, the difference, in fact,  does not  look  very drastic:
just 0.58\,fm instead of 0.66\, fm.  Given that the  hard-wall model
for confinement is  rather  crude, the  agreement may be considered as  encouraging.
Furthermore, one  may  expect that,  in a more realistic
softer model of confinement,  the size of the pion will be
larger. Such an expectation is supported
by our soft-wall model calculation of  the $\rho$-meson
electric radius, for which we obtained
$\langle r^2_{\rho} \rangle_{C}=$0.66\,fm$^2$
 (0.40\,fm$^2$  for  $\langle r^2_{\rho} \rangle_{\cal F}$),
i.e.,  the result by 0.13\,fm$^2$ larger than in the hard-wall model.
If   $\langle r^2_{\pi} \rangle $ would increase
by a  similar amount,  the result
will be   very  close to the quoted  experimental  value.

To   find $\langle r^2_{\pi} \rangle$   for large $a$ (i.e., when $\alpha \gtrsim z_0^{-3}$
 for fixed $z_0$, or when $z_0 \gtrsim \alpha^{-1/3}$ for fixed $\alpha$),
 we use first the observation that,
in the region $a \gtrsim 2$, we may approximate $\nu (a)\approx 1$.
Then the factor in square brackets in Eq.~(\ref{rhozeta}) becomes
a function of the combination $a\zeta^3\equiv \mu$  (call it ${ R} (\mu)$),
and we can write
\begin{align}
\label{rpimu}
 \langle r^2_{\pi} \rangle \Bigl |_{a \gtrsim 2}
\approx  \frac34 \, \Gamma (1/3)\, \Gamma (2/3)\,
 \left ( \frac1{\alpha} \right )^{2/3}\left.  \left. \int_{0}^{a} d\mu
  \,  \mu ^{5/3}\, { R} (\mu) \right [1- \frac23 \, \ln \,\frac{\mu}{a}\right ]    \ .
\end{align}
For $a \gtrsim 2$, the upper limit of integration in this expression
  may be safely substituted
by infinity  producing
\begin{align}
\label{rpim}
   \int_{0}^{\infty} d\mu
  \,  \mu ^{5/3}\, { R} (\mu) =\frac{2^{2/3}}{3 \,  \Gamma^2(2/3)} \equiv G \   \   \   \  ,  \   \   \   \
  \int_{0}^{\infty} d\mu
  \,  \mu ^{5/3}\, { R} (\mu)  \ln \mu \approx G \ln 0.566 \ ,
\end{align}
which gives
\begin{align}
 \langle r^2_{\pi}  \rangle \Bigl |_{a \gtrsim 2}&= \frac{\Gamma(1/3)}{2^{4/3}\Gamma(2/3)}
  \left ( \frac1{\alpha} \right )^{2/3}
\left [ 1+\frac23 \, \ln \left ( \frac{ a} {0.566} \right )   \right ] \  .
\end{align}
Using Eq.~(\ref{fpiinfty}), we   can express the coefficient in front of 
the square bracket in terms of $f_\pi$: 
\begin{align}
\label{rpifpi}
 \langle r^2_{\pi}  \rangle \Bigl |_{a \gtrsim 2}
&=  \frac3{4\pi^2 f_\pi^2} +  \frac1{2\pi^2 f_\pi^2}
 \ln \left ( \frac{ \alpha z_0^3} {0.566} \right )    \  .
\end{align}
Thus, $ \langle r^2_{\pi}  \rangle $   in the  $a\gtrsim 2$ region   consists 
of two componens: a fixed term  $3/4\pi^2 f_\pi^2$
and a term logarithmically increasing with $z_0$.
As $z_0 \to \infty$, the pion charge radius becomes infinite,
reflecting the fact that the pion in this model is  massless.
A similar   structure in the  expression for the pion charge radius  
was obtained   \cite{Hippe:1995hu}  in the Nambu-Jona-Lasinio (NJL)  model 
\begin{align}
 \langle r^2_{\pi}  \rangle _{NJL}
&=  \frac3{2\pi^2 f_\pi^2} +  \frac1{8\pi^2 f_\pi^2}
 \ln \left ( \frac{m_\sigma^2}{m_\pi^2}  \right )    \  .
\end{align}
It also has the  logarithmic term  $\ln m_\pi^2$ \cite{Beg:1973sc,Volkov:1974bi}
resulting in the infinite radius for massless pion  and 
the infrared-finite piece $3/2\pi^2 f_\pi^2$ \cite{Gerasimov:1978cp,Bernard:1988wi}. 
The latter, however, is twice larger than that 
in our result (\ref{rpifpi}) 
and  contributes 0.34\,fm$^2$  to   $ \langle r^2_{\pi}  \rangle $,
with the chiral logarithm term  producing the extra 0.11\,fm$^2$
required for agreement with experiment. 
In our case, the logarithmic term taken for $a=a_0$
is approximately equal to $3/4\pi^2 f_\pi^2$,   thus 
almost  doubling the outcome value for  $ \langle r^2_{\pi}  \rangle $.
More precisely, we can write 
\begin{align}
\label{rpifpi2}
 \langle r^2_{\pi}  \rangle \Bigl |_{a \gtrsim 2}
&=  \frac3{2\pi^2 f_\pi^2}\left [ 1 + \frac13 \, \ln \left (\frac{a}{2.54} \right ) \right ]    \  .
\end{align}
For $a=2.26$, the modified logarithmic term gives a very small contribution,
and our net result is  very  close to the value given by the NJL fixed term.
 Numerically, though,
this  prediction of the hard-wall AdS/QCD model,  as we have seen,  is essentially 
smaller than the 
 experimental value.

\subsection{Form Factor at  Large $Q^2$}

In the large-$Q^2$ limit,   the source ${\cal J} (Q,z)$  is given by
its free-field version
$zQK_1 (Qz)$   that behaves asymptotically like  $e^{-Qz}$.
As a result, only small values  $z\sim 1/Q$   are
important in the form factor integral, and
the large-$Q^2$ asymptotic behavior of the form factor
is determined by the value of $\rho(z)$ at the origin
\cite{Polchinski:2002jw,Brodsky:2003px,Grigoryan:2007my},  namely,
\begin{align}
 F_{\pi}(Q^2) \to \frac{2\,  \rho (0)}{Q^2}= \frac{2\, \phi_\pi^2  (0)}{Q^2} =
\frac{4\pi^2 f_\pi^2}{Q^2} \equiv \frac{s_0}{Q^2} \  .
\end{align}
Note  that the combination   $4\pi^2 f_\pi^2\equiv s_0\approx 0.68\, {\rm GeV}^2$  
frequently appears in the pion studies. In particular, it is the basic 
scale  of the pion wave function 
  in the local quark-hadron duality model \cite{Nesterenko:1982gc,Radyushkin:1995pj}, 
where it corresponds to the  ``pion duality interval''.
\begin{figure}[h]
\mbox{
 {\epsfysize=4cm
 \hspace{0cm}
  \epsffile{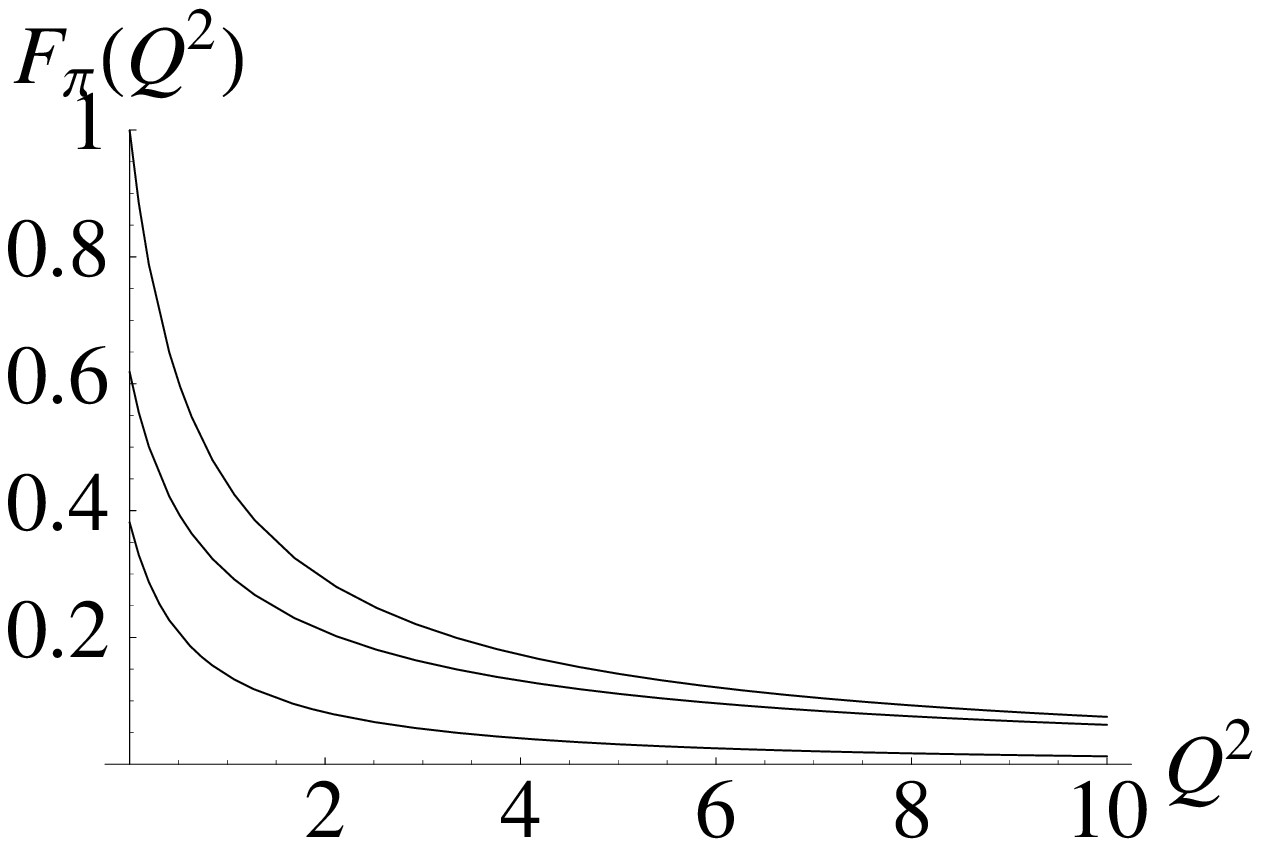} }
  \hspace{1cm}
  {\epsfysize=4cm  \epsffile{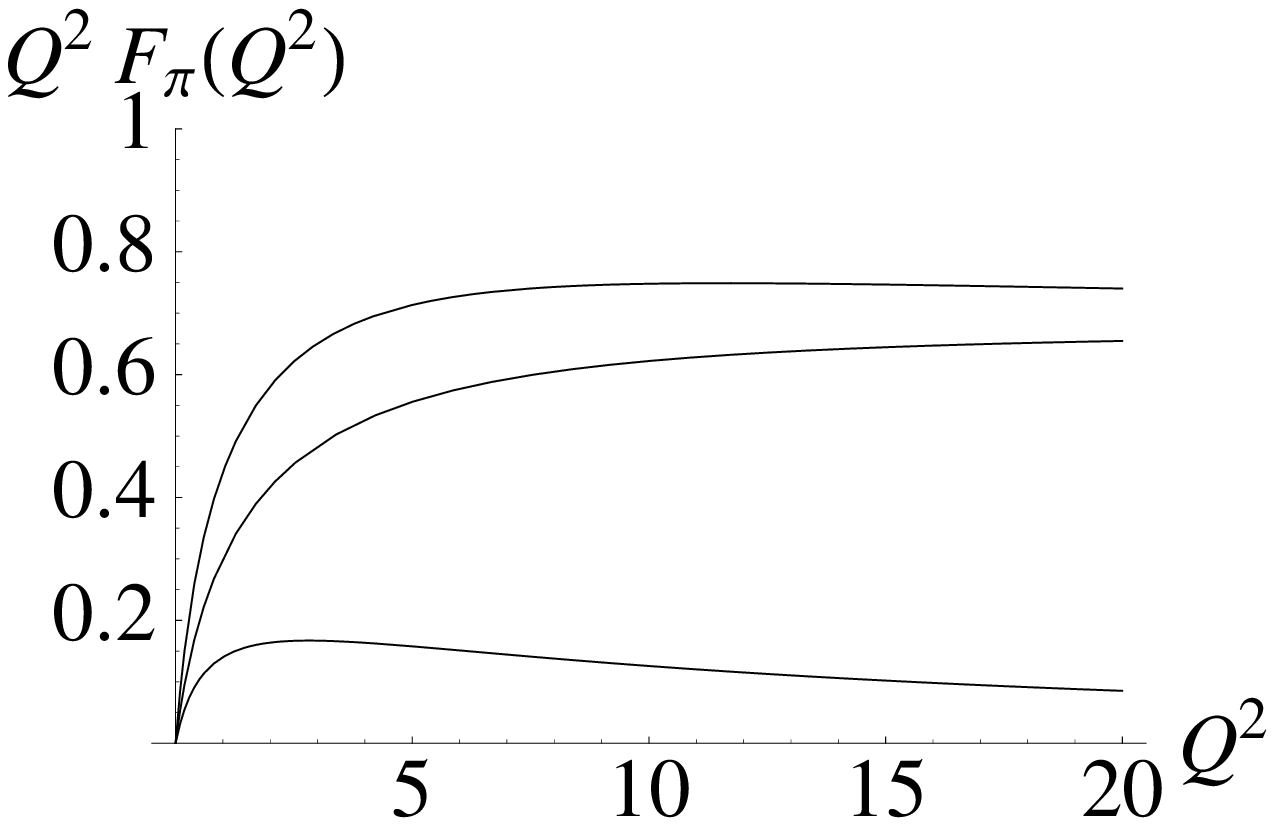}} }\hspace{0cm}
\caption{\label{ffactor}
{\em Left:}  Contributions to pion form factor $F_\pi(Q^2)$ from $\Psi^2$-term (lower curve),
from $\Phi^2$-term (middle curve) and total contribution (upper curve).
{\em Right:} Same for $Q^2 F_\pi (Q^2)$.
}
\end{figure}

The leading contribution comes entirely from the
$\Phi^2$ term of the form factor integral (\ref{FF2}) while the $\Psi^2$ term
contribution behaves asymptotically
like  $1/Q^4$ since it
is accompanied by extra $z^2$ factor.
 Note, however, that it is quite visible in the experimentally
interesting region $Q^2 \lesssim 10\,{\rm GeV}^2$:  it is
 responsible for more than
$20\%$ of the form factor value   in this region (moreover, at $Q^2=0$,
the $\Psi^2$  term contributes about 40\% into the normalization
of the   form factor).

\begin{figure}[h]
\mbox{
 {\epsfysize=4cm
  \epsffile{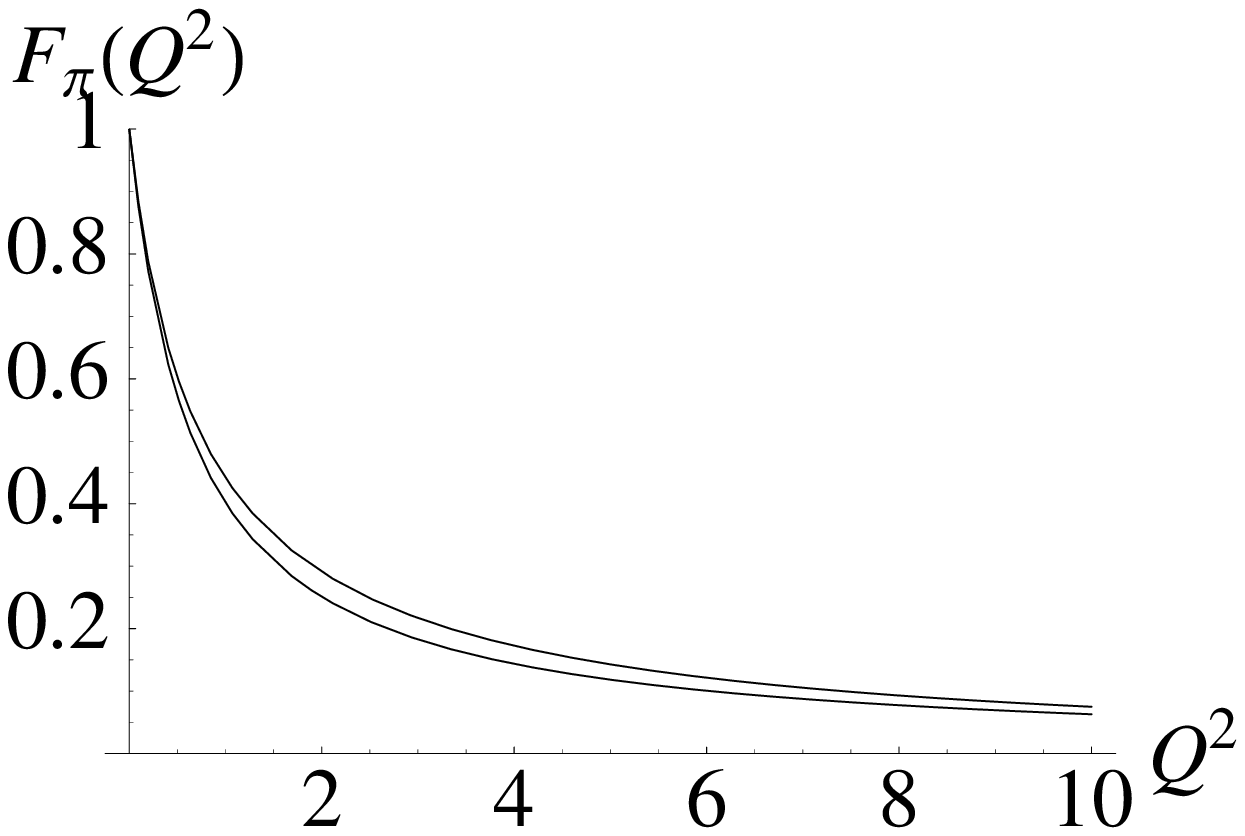} }
  \hspace{1cm}
  {\epsfysize=4cm  \epsffile{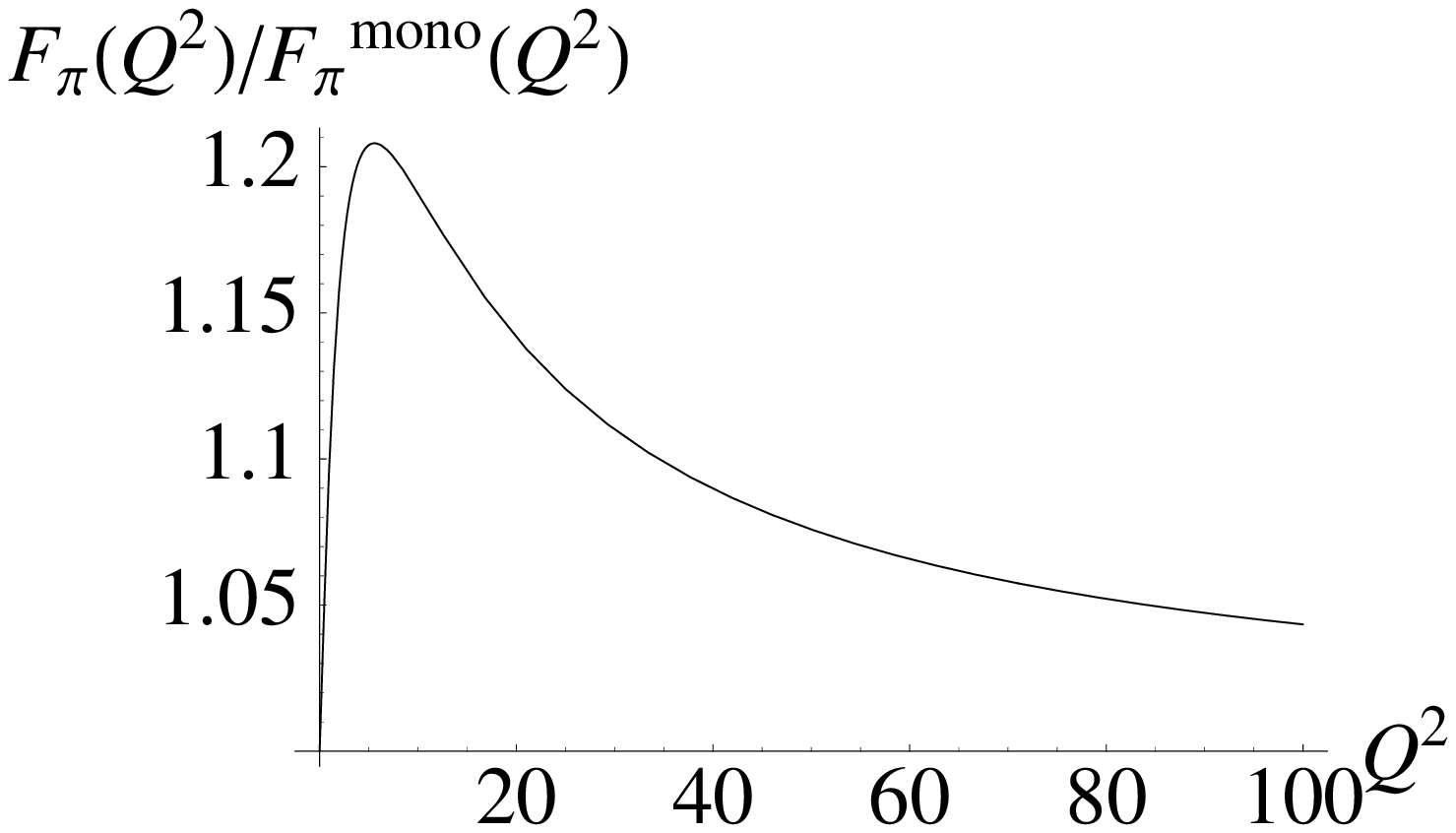}} \hspace{0cm}
 }
\caption{\label{ffact}
{\em Left:} Pion form factor $F_\pi(Q^2)$ from the holographic model (upper curve)
in comparison with the monopole interpolation $F_\pi^{\rm mono}(Q^2)$  (lower curve).
 {\em Right:}  Ratio  $ F_\pi (Q^2)/F_\pi^{\rm mono}(Q^2)$.
}
\end{figure}

From a phenomenological point of view,
different AdS/QCD-like models for the pion
form factor differ in the shape of the  density
$\rho (\zeta)$ that they produce.
If we require  that the  density  $\rho  (z)$
 equals $2 \pi^2 f_\pi^2$ at the origin,
the asymptotic behavior is $F_\pi (Q^2) \to s_0/Q^2$ in
any such model.
For $Q^2=0$, the form factor is normalized to one,
so basically the models would differ in how they interpolate
between these two limits.
 In particular, the simplest  interpolation is provided by the  monopole formula
\begin{align}
 F_\pi^{\rm mono}(Q^2) = \frac1{1+Q^2/s_0}  \   ,
\end{align}
while our hard-wall calculation gives a curve that
goes above $ F_\pi^{\rm mono}(Q^2)$:
the ratio  $F_\pi (Q^2)/F_\pi^{\rm mono}(Q^2) $
is  larger  than 1 for all $Q^2>0$, slowly approaching unity
as $Q^2 \to \infty$ (see Fig.\ref{ffact}).

In fact, a  purely
monopole  form factor
was  obtained in our paper \cite{Grigoryan:2007my}, where we studied
 the $\rho$-meson
form factors in the soft-wall holographic model,
in which confinement is generated by $\sim\!\!z^2$  oscillator-type
potential.
It was shown  in  \cite{Grigoryan:2007my} that
the form factor integral
\begin{align}
\label{calf00}
 {\cal F}  (Q^2, \kappa ) = \int_0^{\infty} dz \, z \, {\cal J}^{\rm O}  (Q,z)\,
|\Phi (z, \kappa)|^2  \,  ,
\end{align}
in which  $\Phi  (z,\kappa) = \sqrt{2} \,\kappa\, e^{- z^2\kappa^2/2}$ is the
lowest bound state wave function, and
\begin{align}
\label{calJO}
 {\cal J}^{\rm O}  (Q,z) = z^2 \kappa^2 \int_0^1 \frac{dx}{(1-x)^2}\,x^{{Q^2}/{4\kappa^2}}
\,  \exp
\left [- \frac{x}{1-x}\, z^2\kappa^2 \right ]
\end{align}
is the bulk-to-boundary propagator of this oscillator-type model,
is  exactly equal to \mbox{$1/(1+Q^2/(4\kappa^2))$.}
The  magnitude  of the oscillator scale $\kappa$ was  fixed in  our paper
\cite{Grigoryan:2007my}
by the value of the $\rho$-meson mass: $\kappa = \kappa_\rho \equiv  m_\rho/2$.
As a result, the form factor ${\cal F} (Q^2, \kappa = m_\rho/2)$
had the $\rho$-dominance
behavior $1/(1+Q^2/m_\rho^2)$.

If we take $\kappa = \kappa_\pi \equiv  \pi f_\pi \approx 410\,{\rm MeV}$
both for $\Phi  (z,\kappa)$ and ${\cal J}^{\rm O}  (Q,z)$,
the integral (\ref{calf00})  gives $1/(1+Q^2/s_0)$.
The relevant wave function $\Phi (z, \kappa_ \pi)$
has the expected  correct normalization
\mbox{$\Phi  (0, \kappa_ \pi) = \sqrt{2} \pi f_{\pi}$,}   however,
the  slope $1/s_0$ of $1/(1+Q^2/s_0)$   at $Q^2=0$  (corresponding to
\mbox{0.35 fm$^2$} for the radius squared) is smaller
than that of the experimental pion form factor.
Furthermore,  $Q^2 F_\pi^{\rm mono}(Q^2) $
 tends to $s_0 \approx 0.68\,$GeV$^2$
for large $Q^2$,  achieving values about $0.5\,$GeV$^2$
for $Q^2 \sim 2\,$GeV$^2$, and  thus exceeding  by   more than 25\%
the experimental JLab values 
\cite{Horn:2006tm}
measured for $Q^2=1.6$ and 2.45\,GeV$^2$.
The authors of Ref.~\cite{Brodsky:2007hb}
 proposed   to use
Eqs.~(\ref{calf00}),(\ref{calJO})  as an AdS/QCD  model
for the  pion form factor, with $\kappa = 375\, $MeV chosen so as
to fit these high-$Q^2$  data.
However, such a choice underestimates the
value of $f_\pi^2$ by almost 30\%.
Our opinion is that  the  AdS/QCD
models should  describe  first  the low-energy properties of hadrons,
and the   basic
low-energy characteristics,  such as $m_\rho$ and $f_\pi$, should be used 
to fix
the model parameters.
On the other hand, if the  form factor calculations based on these
parameters disagree with
 the large-$Q^2$ data, it is quite possible that this  is 
just an indication that one is using the model
beyond its
applicability limits.
Furthermore,  as we have seen in the hard-wall model,
to correctly describe the pion one needs to  
include  the chiral symmetry breaking effects absent in the vector channel.
As a result, equations for pion wave functions 
are rather different from those in the $\rho$-meson case.
Similarly, there are no reasons to expect that,  in a soft-wall model,
the pion density should have the same shape as the $\rho$-meson one.
Unfortunately, 
the procedure of bringing in the chiral
symmetry breaking effects that was used in the hard-wall 
model of Ref.~\cite{Erlich:2005qh} faces serious difficulties
when applied to  the AdS/QCD model \cite{Karch:2006pv} with the $z^2$ soft wall.
   As discussed  in  Ref.~\cite{Karch:2006pv},  the solution
of the equation for the $X$ field in this model requires 
that chiral  condensate $\sigma$  and the  mass parameter $m_q$ are 
proportional to each other, so that $\sigma $ cannot be varied 
independently of $m_q$.  Moreover, if one takes the chiral limit 
$m_q=0$,   the chiral condensate should 
also vanish. 
This  difficulty   may   be   avoided   by  switching to    more sophisticated 
 recent    models   
(cf.  \cite{Casero:2007ae,Gursoy:2007cb,Gursoy:2007er})  
in  which the chiral condensate is   generated dynamically. 
However,   such a consideration goes well beyond the scope of the present paper.
Thus,   we just resort to an idea   that whatever the mechanism is involved,  the net 
practical outcome is a particular shape of the density
$\rho (z)$ that eventually determines the pion form factor
and other pion characteristics. 
Below, we give an example of a  density  $\rho^{\rm mod} ( z)$ that
is  normalized  at the origin  
by  the experimental value of $f_\pi$,  i.e., $\rho^{\rm mod} (0) = 2 \kappa_\pi^2$, but
which is also capable to  reproduce   the experimental value of the pion charge radius.

Evidently,   to   increase the  radius,
 we should take  a density which is  larger for  large $z$ than
$\Phi^2 (z, \kappa = \pi f_\pi)$.
Since the  overall integral normalization  of the  density
is   kept fixed,
this can be achieved only by decreasing
the density for  small $z$ values.

\begin{figure}[h]
\mbox{
  {\epsfysize=4cm  \epsffile{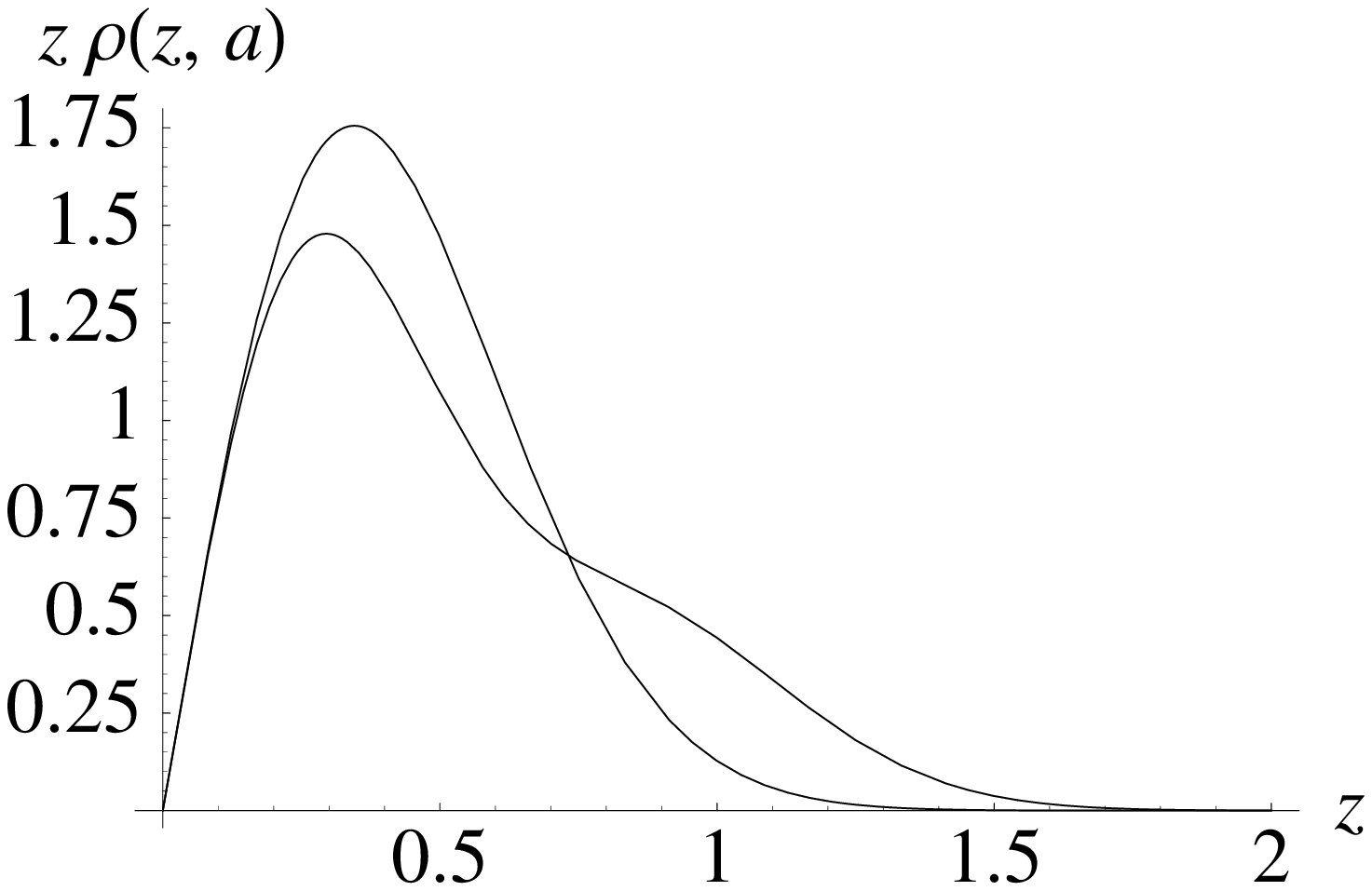}}
\hspace{1cm}
  {\epsfysize=4cm  \epsffile{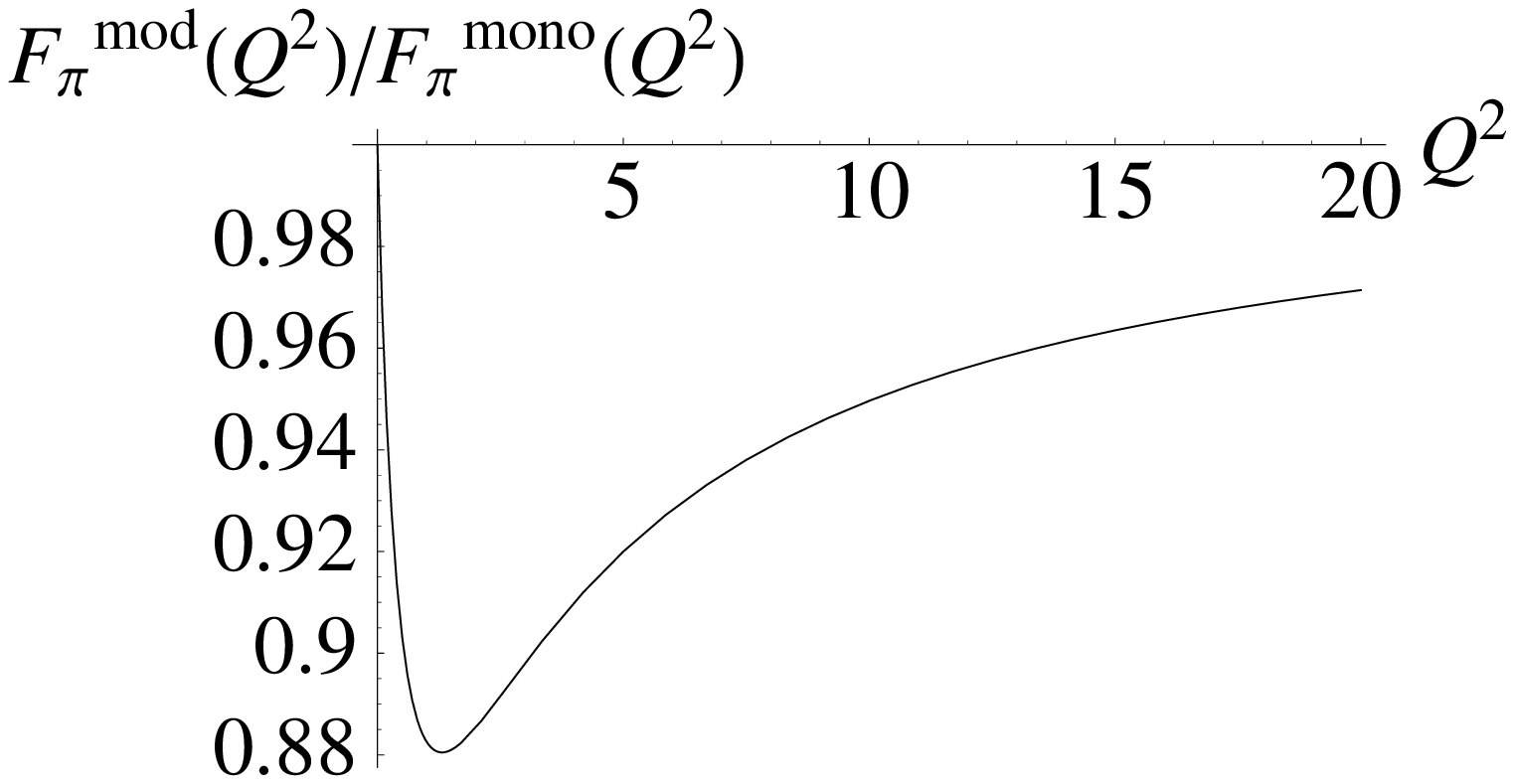}}}
\caption{\label{rhomod}
Left: Model density $z \rho^{\rm mod} (z)$ (measured in fm$^{-1}$)
is larger than the density $z |\Phi (z, \kappa_\pi)|^2$ for large $z$ (displayed in fm).
 Right: Ratio  $ F_\pi^{\rm mod} (Q^2)/F_\pi^{\rm mono}(Q^2)$ for $B=1/4$.
}
\end{figure}

Consider a  simple ansatz (see Fig.\ref{rhomod})
\begin{align}
 \rho^{\rm mod} (z) = 2 \kappa_\pi^2 \, e^{-z^2\kappa_\rho^2} \left [1-A\, z^2\kappa_\rho^2
+B \, z^4\kappa_\rho^4 \right ]  \   ,
\end{align}
with $A= 1- \kappa_\rho^2/\kappa_\pi^2+2B$. It
has both the desired value  for $z=0$ and satisfies
the normalization   condition
\begin{align}
 \int_0^\infty dz \, z   \rho^{\rm mod} (z) =1 \   .
\end{align}
 Integrating it with $ {\cal J}^{\rm O}  (Q,z)$ taken at $\kappa = \kappa_\rho$
produces the model form factor  given by the
following sum of contributions
of the three lowest vector states:
\begin{align}
{F_\pi^{\rm mod} (Q^2)} =
\frac{2-(1-2B)s_0/m_\rho^2}{1+Q^2/m_\rho^2} -
\frac{1-(1-4B)s_0/m_\rho^2}{1+Q^2/2m_\rho^2}+
 \frac{2Bs_0/m_\rho^2} {1+Q^2/3m_\rho^2}\  .
\end{align}
The slope of   $F_\pi^{\rm mod} (Q^2)$  at $Q^2=0$
is given by
\begin{align}
 \frac{dF_\pi^{\rm mod} (Q^2)}{dQ^2} =- \frac1{m_\rho^2}
 \left [ \frac32 -  \left ( \frac 12 - \frac23 B \right )
\frac{s_0}{m_\rho^2}  \right ] \  .
\end{align}
Taking  $B=1/4$, one obtains the experimental value 0.45 fm$^2$ for $\langle r_\pi^2 \rangle$. It is interesting
to note that the model density providing this value, has an   
enhancement for larger values of $z$ (see Fig.\ref{rhomod}), just like the
pion densities in the hard-wall model (see Fig.(\ref{rhodense})). Due to
a  larger slope, $F_\pi^{\rm mod} (Q^2)$
decreases faster than the simple monopole interpolation ${F_\pi^{\rm mono} (Q^2)}$ and,  as a result,   is   in
better agreement with the data. In fact, it  goes very close to $Q^2 \lesssim 1\,$GeV$^2$  data, but exceeds  the
values of the JLab
 $Q^2 =$1.6 and 2.45\,GeV$^2$  points by roughly  10\% and 20\%, respectively.

This  discrepancy has a general  reason. The asymptotic AdS/QCD  prediction  is 
\mbox{$Q^2 F_\pi (Q^2)|_{Q^2 \to \infty} \to 4 \pi^2 f_\pi
^2$} which is $\approx 0.68\,$GeV$^2$ for  experimental value of $f_\pi$. On the other
hand,   JLab  experimental   points correspond to $Q^2 F_\pi^{\rm exp} (Q^2) \approx 0.4\,$GeV$^2$, which is much
smaller than the theoretical  value quoted  above. The pre-asymptotic effects, as we have seen, reduce the
discrepancy, but there still  remains a 
sizable  gap. As we already stated,  such a disagreement may be just a signal that
we are reaching a  region where 
AdS/QCD models should   not be expected to  work. In particular,  AdS/QCD models  of 
Refs.~\cite{Polchinski:2002jw,Brodsky:2003px,Erlich:2005qh,DaRold:2005zs}    
describe  the   pion  in terms of   an effective field or current, without specifying  whether  the
current is built from spin-1/2 fields, or  from scalar fields, etc.
 For $Q^2$ above 1\,GeV$^2$, the quark
substructure of the pion may be resolved by 
the electromagnetic probe (which is a wide-spread belief), and the
description of the pion ``as a whole'' may be insufficient.

\section{Summary} 

In this paper,  we studied  the pion in the chiral limit of two flavor QCD. 
To this end, we described a formalism that allows
to extract pion form factor within the framework of the holographic 
dual model of QCD with hard-wall cutoff.  
 Following Ref.~\cite{Erlich:2005qh}, we   identified the pion 
with the longitudinal component of the axial-vector gauge field.
We defined  two  (Sturm-Liouville) conjugate  wave
functions  $\Phi (z)$ and $\Psi (z)$  that describe the structure of 
the pion   along the  5$^{\rm th}$ dimension 
coordinate $ z $.  These wave functions provide a very convenient 
framework to study the holographic physics of the pion.
We demonstrated that, just like in the $\rho$-meson 
case \cite{Grigoryan:2007vg},   the pion form factor 
is given by an integral involving the function $\rho (z)$ that has 
the meaning of the charge density inside the pion.  However, 
in distinction to  the $\rho$-case,
when the density was simply given by $|\Phi (z)|^2$,
the pion density has an additional term proportional to $|\Psi (z)|^2$ 
and entering with the $z$-dependent coefficient  reflecting 
the mechanism of the spontaneous symmetry breaking.  
Both terms are  required for normalization 
of the form factor at $
Q^2 = 0 $.

We found an  analytic expression for the pion decay constant in terms of two parameters of the model:
$ \sigma $ and $ z_0 $, similar to those used in Ref.~\cite{DaRold:2005zs}. 
Analyzing the results,  we found it  convenient to work with two
combinations  $ \alpha = g_5 \sigma/3 $ and $ a = \alpha z^3_0 $ of the
basic parameters.  In particular,  we found $a=a_0=2.26$ for the value of $a$ corresponding 
to the experimental $\rho$-meson mass $m_\rho$ and pion  decay constant $f_\pi$.
The importance of the parameter $a$ is that its  magnitude 
  determines the regions,  where the pion properties are 
either governed by  the confinement effects or by  the
effects from the spontaneous chiral symmetry breaking. 
For example, in the practically important domain   $ a > 2 $,  
the pion decay constant is determined primarily by   $ \sigma $,
with negligibly tiny corrections due to $z_0$ value. 
However,   when $ a < 1 $  the pion decay constant 
is proportional to the ratio $ a/z_0 $. Besides, for small $ a \ll1 $,
the radius of the pion is given by $ \langle r^2_{\pi} \rangle = \frac{4}{3}z^2_0 $,
i.e., as one may expect,  the pion size is completely 
determined by the  confinement radius.  On the other hand, for $ a > 2 $
the  radius is basically determined by $1/\sigma^{1/3}$,   
slowly   increasing  with  $z_0$ due to the  $ \ln a/a_0 $
correction.

We also found that the pion rms charge  radius $ \langle r^2_{\pi} \rangle^{1/2} \approx 0.58\,$fm
 in the hard-wall  model 
is smaller than that 
measured  experimentally.  In a sense,
the hard wall at the distance $z_0 \approx 0.62\,$fm
 (fixed from the $\rho$-meson mass),
``does not allow'' the pion to get larger.  So, we argued 
 that if   the IR wall  is ``softened'',
the size of the pion may  be  increased by an  amount sufficient
to accomodate the data. 
A straightforward idea is to use the soft-wall model of Ref.~\cite{Karch:2006pv} 
and treat the pion in a way similar to what was  done  in \cite{Grigoryan:2007my}
for the $\rho$-meson case. 
Unfortunately, there are prohibiting complications with   directly introducing the chiral
symmetry effects within the AdS/QCD model with the $z^2$ 
soft wall.  As explained in  Ref.~\cite{Karch:2006pv},  the chiral  condensate $\sigma$
in such a model is proportional to the  mass parameter $m_q$, so that 
in   the chiral limit  the condensate vanishes together with the quark mass. 

To illustrate  a possible  change in the form factor predictions 
due to the 
softening of  the IR wall,
we proposed  an ansatz for the pion density function and used the vector current source from the 
soft-wall model  considered in Ref.~\ \cite{Grigoryan:2007my}. 
We  demonstrated that  this ansatz is capable to 
fit the experimental value of the pion charge radius.
It also closely follows the data in the $Q^2 < 1\,$GeV$^2$ region,
while still overshoots available  data in the $Q^2\sim 2\,$GeV$^2$ region.
The   basic source  of  this discrepancy is very general:  
  the asymptotic AdS/QCD prediction for the pion
form factor is $Q^2 F_\pi (Q^2) \to 4 \pi^2 f_\pi^2$, and  if one takes 
the experimental value for $f_\pi$,  one obtains 
$Q^2 F_\pi (Q^2) \to 0.68\,$GeV$^2$, which is much larger
than the 0.4\,GeV$^2$ value given by $Q^2\sim 2\,$GeV$^2$ JLab data.
For this reason, we argued that the  disagreement mentioned above  may
be a signal that  the region $Q^2 \gtrsim 2\,$GeV$^2$ 
is beyond the applicability region of AdS/QCD models.

Finishing the write-up of this paper,  we have learned that  the paper 
\cite{Kwee:2007dd} addressing the same  problem
was posted into the arxive. We did not observe, however, 
essential overlaps with our ideas
and results.

\section{Acknowledgments} H.G. would like to thank A.~W. Thomas for valuable comments and support at
Jefferson Laboratory, J.~P. Draayer for support at Louisiana State University.

Notice: Authored by Jefferson Science Associates, LLC under U.S. DOE Contract No. DE-AC05-06OR23177. The U.S.
Government retains a non-exclusive, paid-up, irrevocable, world-wide license to publish or reproduce this
manuscript for U.S. Government purposes.


\end{document}